\documentclass[3p,12 pt]{elsarticle}

\usepackage{geometry}
\usepackage{setspace}
\usepackage{mathtools}
\usepackage{amsmath}
\usepackage{amsfonts}
\usepackage{amssymb}
\usepackage{hyperref}
\usepackage{graphicx}
\usepackage{times}
\usepackage{cleveref}
\usepackage{subcaption}
\usepackage[inkscapelatex=false]{svg}
\usepackage{natbib}
\usepackage{amsthm}
\usepackage{epstopdf}
\usepackage{svg}
\usepackage[export]{adjustbox} 
\theoremstyle{definition}

\usepackage[ruled,vlined]{algorithm2e}
\hypersetup{colorlinks=true,linkcolor=blue,linkbordercolor=white,citecolor=blue}

\newcommand*{\defeq}{\mathrel{\vcenter{\baselineskip0.5ex \lineskiplimit 0pt
			\hbox{\scriptsize.}\hbox{\scriptsize.}}}%
	=}

\journal{}
\begin{document}

	\begin{frontmatter}
	
	\title{Constitutive modeling of viscoelastic solids at large strains based on the theory of evolving natural configurations}
	
        \author[mainaddress]{Tarun Singh}
	\author[mainaddress]{Sandipan Paul\corref{correspondingauthor}}
	\ead{sandipan.paul@ce.iitr.ac.in}
	
 	\address[mainaddress]{Department of Civil Engineering, Indian Institute of Technology Roorkee, Roorkee 247667, India}
 \cortext[correspondingauthor]{Corresponding author}
	
\begin{abstract}
   The theory of evolving natural configurations is an effective technique to model dissipative processes. In this paper, we use this theory to revisit nonlinear constitutive models of viscoelastic solids. Particularly, a Maxwell and a Kelvin-Voigt model and their associated standard solids, viz., a Zener and a Poynting-Thompson solids respectively, have been modeled within a Lagrangian framework. We show that while a strain-space formulation of the evolving natural configurations is useful in modeling Maxwell-type materials, a stress-space formulation that incorporates a rate of dissipation function in terms of the relevant configurational forces is required for modeling the Kelvin-Voigt type materials. Furthermore, we also show that the basic Maxwell and Kelvin-Voigt models can be obtained as limiting cases from the derived standard solid models. Integration algorithms for the proposed models have been developed and numerical solutions for a relevant boundary value problem are obtained. The response of the developed models have been compared and benchmarked with experimental data. Specifically, the response of the novel Poynting-Thompson model is studied in details. This model shows a very good match with the existing experimental data obtained from a uniaxial stretching of polymers over a large extent of strain. The relaxation behavior and rate effects for the developed models have been studied.
\end{abstract}
	
	
\end{frontmatter}

\section{Introduction}

Rheological models are often used to describe the response of viscoelastic materials within a small-strain theory. These models typically consist of two types of basic rheological elements: a spring, representing the non-dissipative elastic response, and a dashpot accounting for the rate-dependent, dissipative response of the material. Different arrangements of these elements either in a series or a parallel connection have long been used to model the response of different aspects of viscoelastic materials such as stress relaxation, creep, fading memory effects etc. These models, however, are developed within a small strain theory which precludes their use in a large class of materials. For example, soft elastomers and biological tissues often exhibit viscoelastic response, but their deformation often goes well beyond the small strain limit. Recent advances in the extensive use of these materials have prompted a need for the extension of these rheological models to a thermodynamically consistent finite deformation framework.

Recognizing this need, many attempts have been made to develop thermodynamically consistent nonlinear viscoelastic models. In particular, finite deformation versions of a Maxwell model~\cite{haupt2002finite}, its generalized form~\cite{govindjee1992mullins} and the associated standard solid model~\cite{lubliner1985model,holzapfel1996large,zhou2018micro,feng2021finite} garnered much attention. To develop these models, a multiplicative decomposition of the deformation gradient~\cite{bilby1955,kroner1959,lee1969} is typically used. In conjunction with an additive split of the Helmholtz free energy its equilibrium and non-equilibrium parts. Thereafter, the evolution equation is derived by requiring the resulting dissipation to be non-negative. These models have been formulated from the current~\cite{reese1998theory}, intermediate~\cite{lion1997physically} as well as the reference configuration~\cite{le1993three}~(see~\cite{gouhier_comparison_2024} for a comparison between these formulations). Using the same modeling strategy, the problem was also recast into a framework involving two potential functions, viz., a Helmholtz free energy accounting for the energy stored in the material and a dissipation potential~\cite{kumar2016two,sadik2024nonlinear}.

While Maxwell-type models have been widely used and implemented into commercial finite element packages, their Kelvin-Voigt counterparts have received considerably less attention. This is due to the fact that unlike the Maxwell-type models, a Kelvin-Voigt model does not naturally fit into the framework of a multiplicative decomposition of the deformation gradient. Nevertheless, a Kelvin-Voigt type standard solid (Poynting Thompson) model has been developed by Huber and Tsakmakis~(2000)~\cite{huber2000finite} by extending the idea of additive decomposition of the Helmholtz free energy. An alternative framework for Kelvin-Voigt model adopts an additive split of the Cauchy or Piola-Kirchhoff stress tensor instead of the multiplicative decomposition of the deformation gradient. This modeling technique is evocative of Simo's~(1987)~\cite{simo1987fully} overstress model which has been widely used in finite viscoelasticity~\cite{kaliske1997formulation,holzapfel2002structural}. Although this modeling technique has been extensively adopted and made its way into commercial finite element (FE) packages, its thermodynamic consistency is still debated~\cite{govindjee2014dynamic,liu2021continuum}. A possible explanation for the veracity of this model was provided by Rajagopal~(2009)~\cite{rajagopal2009note} where he considered the Kelvin-Voigt model as a mixture between an elastic solid and a Newtonian fluid with equal volume fractions. The constitutive relation for the Kelvin-Voigt model, however, was obtained based on this rationale without any formal derivation. Based on this idea, different variants of Kelvin-Voigt model have been proposed particularly in the context of implicit constitutive theories~\cite{coco2023kelvin,itou2023generalization,de2025extensions}. 

Recently Singh and Paul~(2025)~\cite{singh2025} pointed out that the issues of thermodynamic inconsistency is not limited to a Kelvin-Voigt model, but persists in any rheological model associated with a parallel connection within a large deformation setting. A potential remedy for this problem was proposed using the theory of evolving natural configurations proposed by Eckart~\cite{eckart1948thermodynamics} and extensively developed by Rajagopal~(1995)~\cite{rajagopal1995multiple}. In this theory, it is assumed that a body can have multiple natural configurations that are obtained through a local elastic unloading from the current configuration of the body. Akin to~\cite{kumar2016two,sadik2024nonlinear}, this theory also requires specification of the Helmholtz free energy, $\psi$ and a rate of dissipation function, $\xi$. While the former accounts for the elastic response of the body measured from a fixed natural configuration, the latter governs the evolution of the natural configurations. The primary advantage of this theory lies in the determination of the evolution equation. Unlike the traditional Coleman-Noll procedure, here a only non-negative values of $\xi$ are stipulated as admissible which ensures that a Clausius-Duhem form of the second law of thermodynamics is identically satisfied. Thereafter, the evolution equation is determined through a constrained maximization of the rate of dissipation function~\cite{rajagopal2004thermomechanics,ziegler1963}. This treatment allows one to choose the rate of dissipation function either in terms of the kinematic variables or the configurational forces responsible for the evolution of the natural configurations. We refer to these techniques as a strain-space and stress-space formulation, respectively. The strain-space formulation has previously been used in obtaining the finite deformation versions of several viscoelastic fluid models such as a Maxwell and an Oldroyd model~\cite{rajagopal2001modeling}, a Burgers model~\cite{malek2015variant}, etc. While the stress-space formulation is shown to be effective in the context of plasticity to retrieve a $J_2$ plasticity~\cite{rajagopal1998mechanics,rajagopal2004thermomechanics} or a Drucker-Prager model~\cite{paul2021constitutive,paul2025}, its use in finite viscoelasticity has been rather limited. Using this framework, Singh and Paul~\cite{singh2025} proposed a modeling strategy in which the distribution of mechanical power is stipulated to be in the ratio of the norm of the strain rates or the configurational forces for a series or a parallel connection, respectively. Based on this idea, we develop thermodynamically consistent constitutive models for viscoelastic materials undergoing large deformation. The main departure from the existing models is the derivation of the constitutive relations for the Kelvin-Voigt type solids. We show that when the rate of dissipation function is written as a quadratic potential of the strain rates, a Maxwell-type model is derived whereas choosing $\xi$ to be a function of the configurational forces leads to a Kelvin-Voigt type material model. Furthermore, we derive the Kelvin-Voigt model as a limiting case of the derived Maxwell-type standard solid (Zener) model.

The rest of the paper is organized as follows. In~\S~\ref{sec:Prelimilaries}, we briefly review the theory of an evolving natural configurations. Particularly, a strain-space formulation of this theory and its applications to Maxwell and the associated standard solid model have been briefly revisited. Here we derive the models within a Lagrangian framework as opposed to the Eulerian formulation of the rate-type fluids of Rajagopal and Srinivasa~(2001)~\cite{rajagopal2001modeling}. In~\S~\ref{sec:Configurational_force}, we first derive the required configurational force (Eshelby energy-momentum tensor) for a chosen functional form for the Helmholtz free energy. Based on this configurational force, a set of general constitutive relations for compressible and incompressible materials is formulated. A constitutive model for Poynting-Thompson solids is then derived from this formulation in~\S~\ref{sec:applications_stress_space}. In ~\S~\ref{sec:limiting}, the finite deformation version of a Kelvin-Voigt model is derived as a limiting case of the Maxwell-type solids. To demonstrate the efficacy of the developed models, we develop numerical solutions for relevant boundary value problems in ~\S~\ref{sec:numerical_example} and verify the proposed models with available experimental data. Finally, the paper is concluded with summarization of the present work and possible future directions.  

\section{Theory of an evolving natural configurations: Strain space formulation}\label{sec:Prelimilaries}

We briefly revisit the theory of evolving natural configurations developed by Rajagopal~\cite{rajagopal1995multiple} and Rajagopal and Srinivasa~(1998)~\cite{rajagopal1998mechanics}, which serves as an essential ingredient in our proposed models. In this section, we only highlight the strain-space formulation and its application to derive Maxwell-type models.    

\subsection{Kinematics}

Let us consider a simply connected homogeneous body $\boldsymbol{\mathcal{B}}$ embedded in an Euclidean point space. At time $t=0$, the embedding of the body is identified with a subset of this point space, called the undeformed configuration of the body, $\kappa_r(\mathcal{B})$. Let us denote the position vector of a material particle in $\kappa_r(\boldsymbol{\mathcal{B}})$  by $\mathbf{X}$. A motion of the body is a smooth bijective map $\mathbf{x}\defeq\boldsymbol{\mathcal{X}}(\mathbf{X},t)$ that maps it to the position vector of the same particle in its current configuration $\kappa_t(\boldsymbol{\mathcal{B}})$ at time $t$, denoted by $\mathbf{x}$. Let $d\mathbf{X}$ and $d\mathbf{x}$ denote an infinitesimal fiber in the undeformed and current configuration of the body, respectively. The deformation gradient, $\mathbf{F}$, is a tangent map that takes the infinitesimal fiber $d\mathbf{X}$ from the tangent space of $\kappa_r(\boldsymbol{\mathcal{B}})$ and places it into the tangent space of $\kappa_t(\boldsymbol{\mathcal{B}})$, i.e., $d\mathbf{x}=\mathbf{F}d\mathbf{X}$. We further require that the Jacobian of the map $J=\text{det}(\mathbf{F})>0$. In terms of the motion, the deformation gradient $\mathbf{F}$ can be written as
\begin{equation}\label{eq:deformation_gradient}
     \mathbf{F}=\nabla_{\mathbf{X}}\,\boldsymbol{\mathcal{X}}(\mathbf{X},t).
\end{equation}
We also write the velocity associated with this motion as a material time-derivative of the motion and the velocity gradient, $\mathbf{L}$ as 
\begin{equation}\label{eq:velocity_gradient}
    \mathbf{v}\defeq\dfrac{D\,\mathbf{x}(\mathbf{X,t})}{{Dt}},\quad\text{and}\quad\mathbf{L}\defeq\nabla_\mathbf{x}\,\mathbf{v} = \boldsymbol{\dot{\mathbf{F}}}\,\mathbf{F}^{-1}.
\end{equation}
The velocity gradient is decomposed into its symmetric and skew-symmetric parts, as
 \begin{equation}
     \mathbf{D}\defeq\dfrac{1}{2}\left(\mathbf{L}+\mathbf{L}^T\right)\quad\text{and}\quad\mathbf{W}\defeq\dfrac{1}{2}\left(\mathbf{L}-\mathbf{L}^T\right).
 \end{equation}
Here $\mathbf{D}$ is called a rate of deformation tensor and $\mathbf{W}$ is the spin tensor.
The right and left Cauchy Green tensor, denoted by $\mathbf{C}$ and $\mathbf{B}$, respectively, are defined as
\begin{equation}\label{eq:Cauchy-Green_tensors}
\mathbf{C}\defeq\mathbf{F}^T\mathbf{F},\quad\text{and}\quad\mathbf{B}\defeq\mathbf{F}\mathbf{F}^T.
\end{equation}
These two tensors play an important role as they relate the squared length of an infinitesimal fiber in the undeformed and current configuration of the body. Specifically, the squared length of the infinitesimal fiber in $\kappa_t(\boldsymbol{\mathcal{B}})$ is written as $d\,\mathbf{x}\cdot d\,\mathbf{x}=d\,\mathbf{X}\cdot\mathbf{C}\,d\,\mathbf{X}$. Similarly, the squared length of an infinitesimal fiber in $\kappa_r(\boldsymbol{\mathcal{B}})$ can be written in terms of $d\mathbf{x}$ as $d\mathbf{X}\cdot d\mathbf{X}=d\mathbf{x}\cdot\mathbf{B}^{-1}\,d\mathbf{x}$. It is important to note that the right Cauchy-Green tensor $\mathbf{C}$ serves as a metric of the undeformed configuration of the body, $\kappa_r(\boldsymbol{\mathcal{B}})$. Now, it is possible to define strain tensors based on these Cauchy-Green tensors.  The Green strain is defined as 
\begin{equation}\label{eq:Green-Lagrangian strain}
    \mathbf{E}\defeq\dfrac{1}{2}\left(\mathbf{C}-\mathbf{I}\right).
\end{equation}
 The rate of deformation tensor can be pulled back into the undeformed configuration $\kappa_r(\boldsymbol{\mathcal{B}})$ to produce the material derivative of the Green strain through 
\begin{equation}\label{eq:Green_strain_rate}
    \mathbf{\dot{E}}=\mathbf{F}^T\,\mathbf{D}\,\mathbf{F}.
\end{equation}

\subsection{Evolving natural configurations}\label{sec:Multiple_natural_configurations}

\begin{figure}[h]
    \centering
    {\includegraphics[width=0.6\columnwidth]{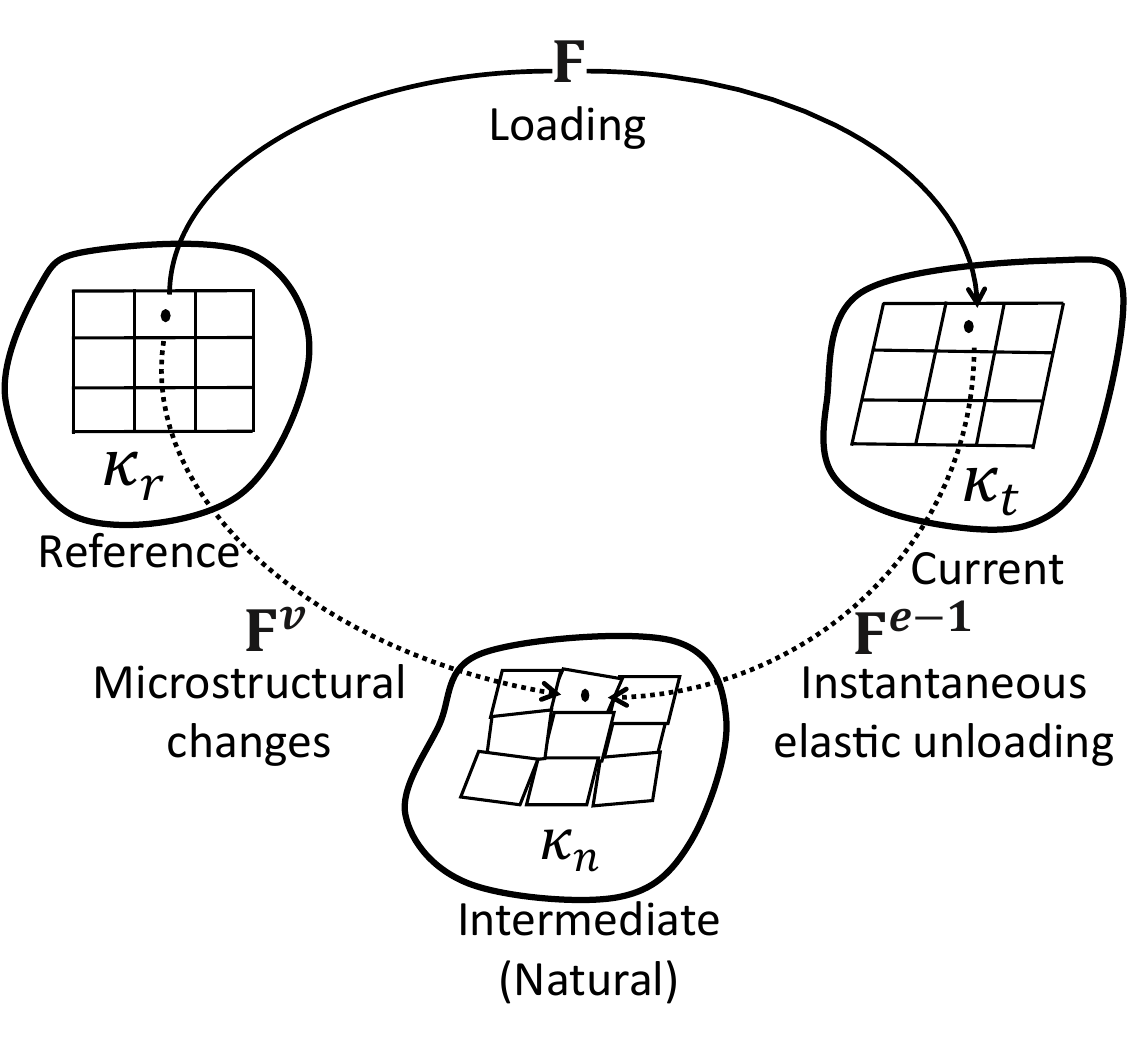}}
    \caption{Evolving natural configurations with the associated tangent maps.}
    \label{fig:Two_term decomposition}
\end{figure}

To model the viscoelastic behavior of materials, the total deformation gradient $\mathbf{F}$ is multiplicatively decomposed into its elastic ($\mathbf{F}^e$) and viscous ($\mathbf{F}^v$) parts as
\begin{equation}\label{eq:F=FeFv}
      \mathbf{F}=\mathbf{F}^e\,\mathbf{F}^v.
  \end{equation}
The physical interpretations of these tangent maps have been discussed in details by Rajagopal and Srinivasa~\cite{rajagopal1998mechanics,rajagopal2004thermomechanics}. According to their theory, at a particular time instant, if an infinitesimal neighborhood of a particle in the current configuration of the body is instantaneously elastically unloaded through the tangent map $\mathbf{F}^{e^{-1}}$ (i.e., all the surface tractions are instantaneously removed), the body occupies a local natural configuration, denoted by $\kappa_n(\boldsymbol{\mathcal{B}})$. In general, the natural configuration of the body is not globally compatible as shown in Fig.~\ref{fig:Two_term decomposition}. Therefore, unlike the total deformation gradient $\mathbf{F}$, the tangent maps $\mathbf{F}^e$ and $\mathbf{F}^v$ generally cannot be obtained as a gradient of a deformation map. It should be noted here that unlike anelasticity, the local natural configuration is not stress-free. In fact, the local natural configuration is subjected to residual stresses and the body occupies only a stress-free configuration upon further viscous relaxation. Therefore, a viscoelastic material can be viewed as a body that has \emph{evolving} natural configurations. 

Now from the elastic and viscous parts of the deformation gradient, the associated right and left Cauchy-Green stretch tensors can be defined as
\begin{equation}
    \mathbf{C}^e\defeq\mathbf{F}^{e^T}\mathbf{F}^e\quad \mathbf{C}^v\defeq\mathbf{F}^{v^T}\mathbf{F}^v\quad \mathbf{B}^e\defeq\mathbf{F}^{e}\mathbf{F}^{e^T}\quad\mathbf{B}^v\defeq\mathbf{F}^{v^T}\mathbf{F}^{v^T}.
\end{equation}
For our analysis, it will be useful to define the kinematic variables in terms of strains similar to the Green strain tensor, $\mathbf{E}$. We define the strain tensors corresponding to the elastic and viscous parts of the deformation gradient as
\begin{align}\label{eq:Ee_Ev}
     \mathbf{E}^e\defeq\dfrac{1}{2}\left(\mathbf{C}^e-\mathbf{I}\right),\quad\text{and}\quad \mathbf{E}^v\defeq\dfrac{1}{2}\left(\mathbf{C}^v-\mathbf{I}\right).
\end{align}
A routine calculation yields the relationship between the total strains and their elastic and viscous counterparts as  
\begin{equation}\label{eq:Ee_Ev_and*}
\mathbf{E}=\mathbf{E}^v+\mathbf{F}^{v^{T}}\mathbf{E}^e\mathbf{F}^v.
\end{equation}
Let us also define the corresponding velocity gradients as
\begin{equation}\label{eq:Le_Lv}
\quad\mathbf{L}^e\defeq\boldsymbol{\dot{\mathbf{F}^e}}\,\mathbf{F}^{e^{-1}}=\mathbf{D}^e+\mathbf{W}^e,\quad\text{and}\quad \mathbf{L}^v\defeq\boldsymbol{\dot{\mathbf{F}^v}}\,\mathbf{F}^{v^{-1}}=\mathbf{D}^v+\mathbf{W}^v.
\end{equation}
Using Eq.~\eqref{eq:F=FeFv}, the total velocity gradient can be written in terms of its elastic and viscous parts as  
\begin{equation}\label{eq:L_LeLv}
    \mathbf{L}=\mathbf{L}^e+\mathbf{F}^e\,\mathbf{L}^v\,\mathbf{F}^{e^{-1}}.
\end{equation}

\subsection{Constitutive relations}\label{sec:constitutive_relation}
In the theory of evolving natural configurations, constitutive relations are derived by specifying two thermodynamic quantities-- a Helmholtz free energy, $\psi$ and a rate of dissipation function, $\xi$. From the first law of thermodynamics, we obtain the relationship between these two quantities as  
\begin{equation}\label{eq:total_work_done}
    \xi=\mathbf{S}\boldsymbol{:}\mathbf{\dot{E}}-\rho_0\,\dot{\psi}\ge0
\end{equation}
where $\mathbf{S}$ is the second Piola Kirchhoff stress. Now let us assume the functional forms for the Helmholtz free energy and the rate of dissipation. Since the Helmholtz free energy $\psi$ accounts for the elastic response of the body measured from a fixed natural configuration, it is reasonable to assume $\psi$ as a function of the elastic Green strain tensor $\mathbf{E}^e$ or alternatively, in view of Eq.~\eqref{eq:Ee_Ev_and*}, it can be written as
\begin{equation}\label{eq:psi}
    \psi=\overline{\psi}\left(\mathbf{E}, \mathbf{E}^v\right).
\end{equation}
On the other hand, the rate of dissipation function accounts for the evolution of the natural configuration. Therefore, its functional form is assumed as
\begin{equation}\label{eq:xi}
    \xi=\overline{\xi}\left(\mathbf{E}^v, \mathbf{\dot{E}}^v\right)\quad\text{with}~\,\overline{\xi}\left(\mathbf{E}^v, \mathbf{0}\right)=0.
\end{equation}
The last part of Eq.~\eqref{eq:xi} implies that whenever there is no change in the natural configuration, i.e., $\mathbf{\dot{E}}^v=0$, then the rate of dissipation should also be zero resulting in a purely elastic process. We further assume that the elastic response of the body for a given natural configuration is that of a Green elastic solid such that
\begin{equation}\label{eq:green_elastic_solid}
    \mathbf{S}=\rho_0\dfrac{\partial\psi}{\partial\mathbf{E}}.
\end{equation}
Using Eq.~\eqref{eq:green_elastic_solid} and the functional forms of $\psi$ and $\xi$ in Eqs.~\eqref{eq:psi} and~\eqref{eq:xi} into the balance of energy Equation.~\eqref{eq:total_work_done}, we obtain a thermodynamic constraint for the rate of dissipative function. Which reads
\begin{equation}\label{eq:xi_constraint}
    \xi=-\rho_0\dfrac{\partial\overline{\psi}}{\partial\mathbf{E}^v}\boldsymbol{:}\mathbf{\dot{E}}^v.
\end{equation}
The evolution equation for $\mathbf{E}^v$ is now derived following the procedure given by Rajagopal and Srinivasa~(2004)~\cite{rajagopal2004thermomechanics}. In a traditional Coleman-Noll procedure~\cite{noll1967}, an expression for the rate of dissipation is obtained from the balance of energy and thereafter, the evolution equation for the internal variable such as $\mathbf{E}^v$ is obtained in such a way that the non-negativity of the rate of dissipation is ensured. Rajagopal and Srinivasa~(2004)~\cite{rajagopal2004thermomechanics} adopted an alternative approach where only those kinematic variables that render the rate of dissipation function non-negative are considered to be admissible. The evolution equation is then obtained by employing a criterion of maximum entropy production. This procedure poses the derivation of the evolution equation as a constrained optimization problem where the rate of dissipation is maximized subject to the constraint~\eqref{eq:xi_constraint}. This departure has a major consequence in our derivation of a Kelvin-Voigt type material model which will be discussed later. Following this procedure, the evolution equation of $\mathbf{E}^v$ for a compressible material can be derived via a Lagrange multiplier technique as
\begin{equation}\label{eq:evoluation_eqn_E_v}
    \dfrac{\partial\overline{\xi}}{\partial{\mathbf{\dot{E}}}^v}=-\dfrac{\lambda}{\left(1+\lambda\right)}\rho_0\dfrac{\partial\overline{\psi}}{\partial\mathbf{E}^v}
\end{equation}
where $\lambda$ is a Lagrange multiplier which can be obtained by the satisfaction of the constraint equation~\eqref{eq:xi_constraint}. 

For an incompressible material model, additional constraints associated with the change in mass density (or volume) must be incorporated. In a Lagrangian framework, this constraint is written as $\text{det}(\mathbf{F})=J=1$. In addition to the total volume, it is reasonable to assume that there will be no change in the volume during the viscous relaxation process, i.e., $\text{det}(\mathbf{F}^v)=J^v=1$. These two constraints enter into the constitutive relations through a modification of the Helmholtz free energy into $\psi_{inc}=\psi-p~(J-1)-q~(J^v-1)$ where $p(\mathbf{X},t)$ and $q(\mathbf{X},t)$ are Lagrange multipliers~\cite{sadik2024nonlinear}. Therefore, from Eq.~\eqref{eq:green_elastic_solid}, the second Piola-Kirchhoff stress can be written as
\begin{equation}\label{eq:green_inc}
    \mathbf{S}=\rho_0\dfrac{\partial\psi}{\partial\mathbf{E}}-p\,J\,\mathbf{C}^{-1}.
\end{equation}
Since the rate of dissipation is a function of $\mathbf{E}^v$ and $\mathbf{\dot{E}}^v$, it needs to be maximized with an additional constraint of $J^v-1=0$. Therefore, the Lagrangian for maximizing the rate of dissipation reads
\begin{equation}\label{eq:_L_inc}
    \mathcal{L}_{inc}=\xi+\lambda\,\left[\xi+\left(\rho_0\,\dfrac{\partial\psi}{\partial\mathbf{E}^v}-q\,J^v\,C^{v^{-1}}\right)\boldsymbol{:}\mathbf{\dot{E}}^v\right].
\end{equation}
Now maximizing the Lagrangian in Eq.~\eqref{eq:_L_inc} with respect to $\mathbf{\dot{E}}^v$ and noting that $J^v$ is a function of $\mathbf{E}^v$ only, the evolution equation~\eqref{eq:evoluation_eqn_E_v} takes on a new form as
\begin{equation}\label{eq:evoluation_eqn_E_v_InC}
    \dfrac{\partial\overline{\xi}}{\partial\dot{\mathbf{E}}^v}=-\dfrac{\lambda}{\left(1+\lambda\right)}\left(\rho_0\,\dfrac{\partial\overline{\psi}}{\partial\mathbf{E}^v}-q\,J^v\,\mathbf{C}^{v^{-1}}\right).
\end{equation}
One can easily notice the similarity between our evolution equation and the kinetic equation for incompressible materials derived by le Tallec~\textit{et al.}~(1993)~\cite{le1993three} and Sadik and Yavari~(2024)~\cite{sadik2024nonlinear}.

\subsection{Modeling of Maxwell-type materials}\label{Sec:Maxwell_type_material}  
In this section, we show the utility of the chosen framework by revisiting some of the well-established material models in finite viscoelasticity. As mentioned earlier, the theory of evolving natural configurations have been previously used to derive constitutive models for rate-type fluids that include Maxwell, Oldroyd-B and Burgers' model~\cite{rajagopal2001modeling,malek2015variant}. We are interested in developing constitutive models for viscoelastic solids within this framework. Specifically, a Lagrangian formulation of the Maxwell model as well as its associated standard solid (Zener) model have been developed. These models are used for benchmarking our theory with the other prior works and their derivations pave the way for further development.
\begin{figure}[tbp]
	\centering
	\begin{subfigure}{0.44\linewidth}
		\includegraphics[width=\columnwidth]{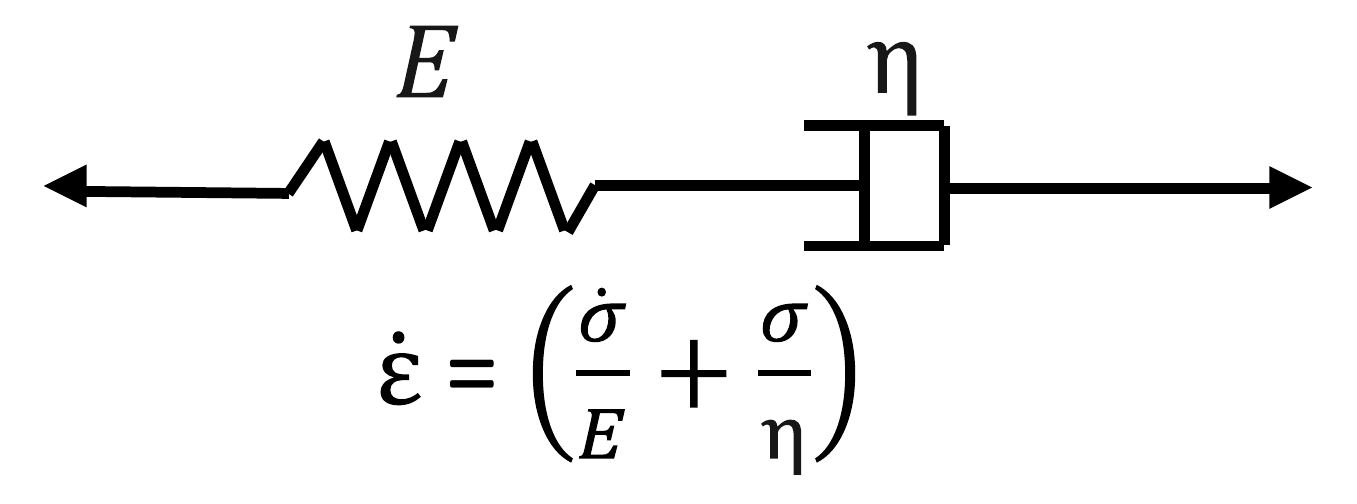}
		\caption{}
		\label{fig:MM}
	\end{subfigure}\hspace{50 pt}
	\begin{subfigure}{0.44\linewidth}
		\includegraphics[width=\columnwidth]{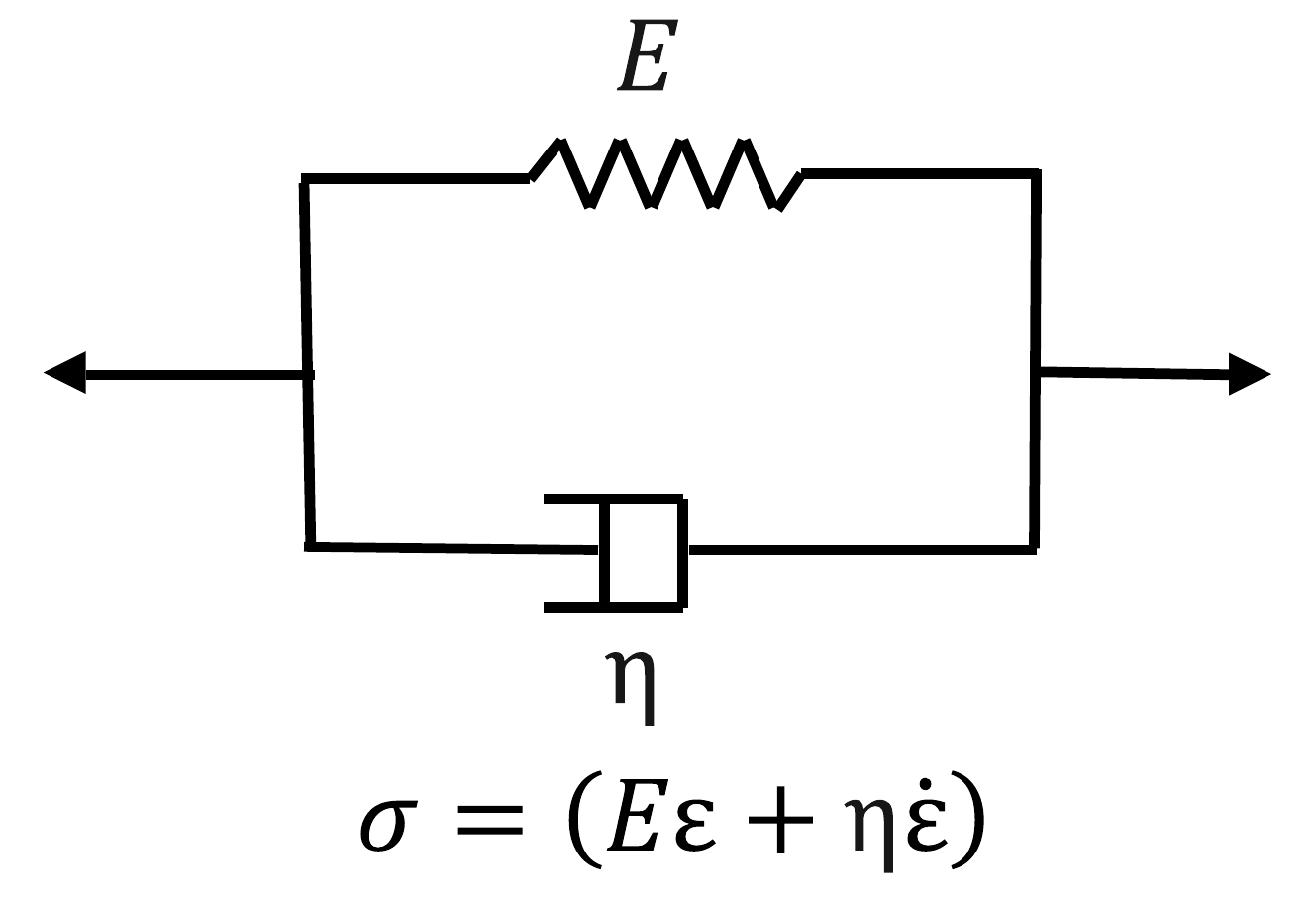}
		\caption{}
		\label{fig:KV}
	\end{subfigure}\\
	\begin{subfigure}{0.44\linewidth}
		\includegraphics[width=\columnwidth]{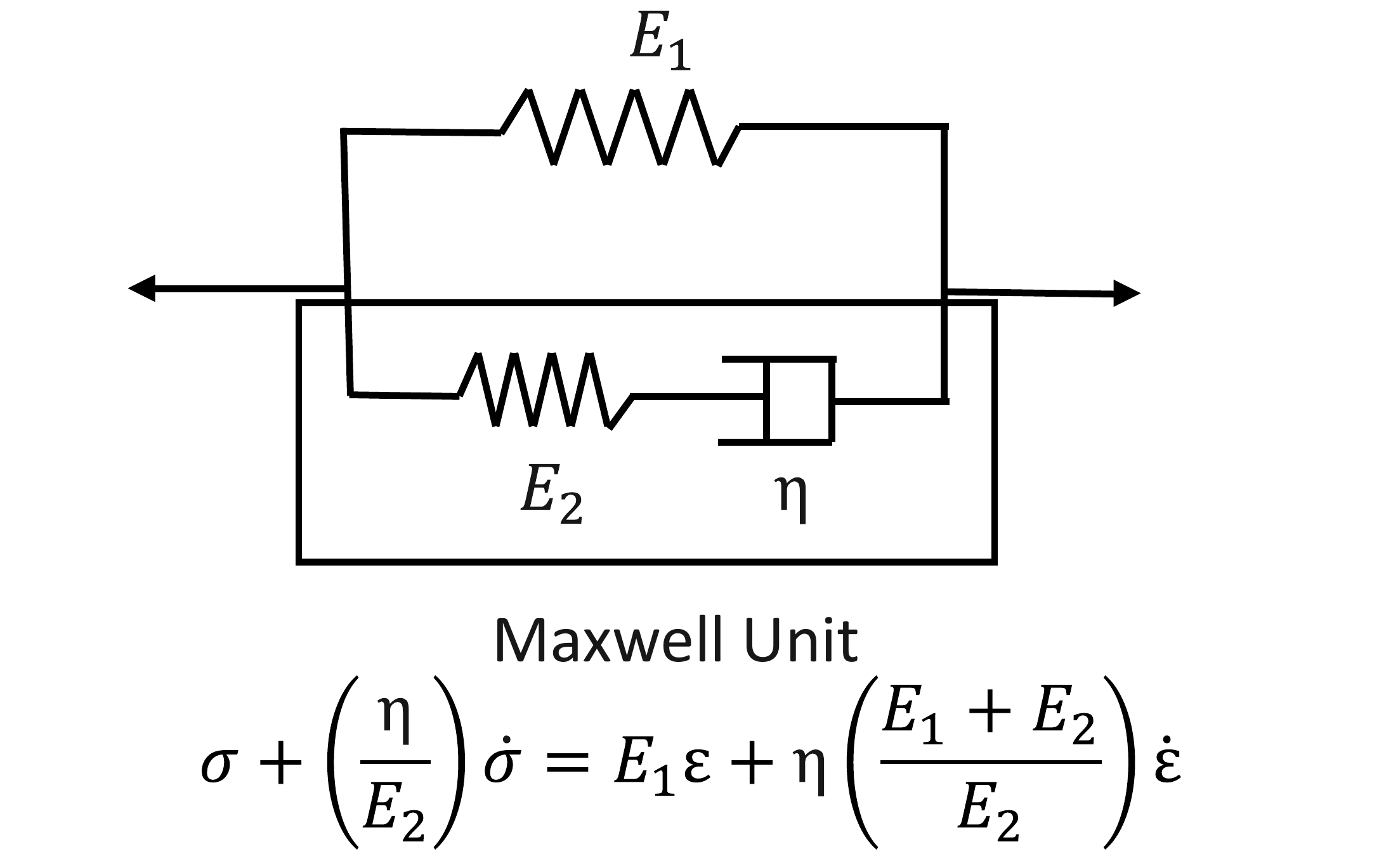}
		\caption{}
		\label{fig:SLS_1}
	\end{subfigure}\hspace{50 pt}
	\begin{subfigure}{0.44\linewidth}
	\includegraphics[width=\columnwidth]{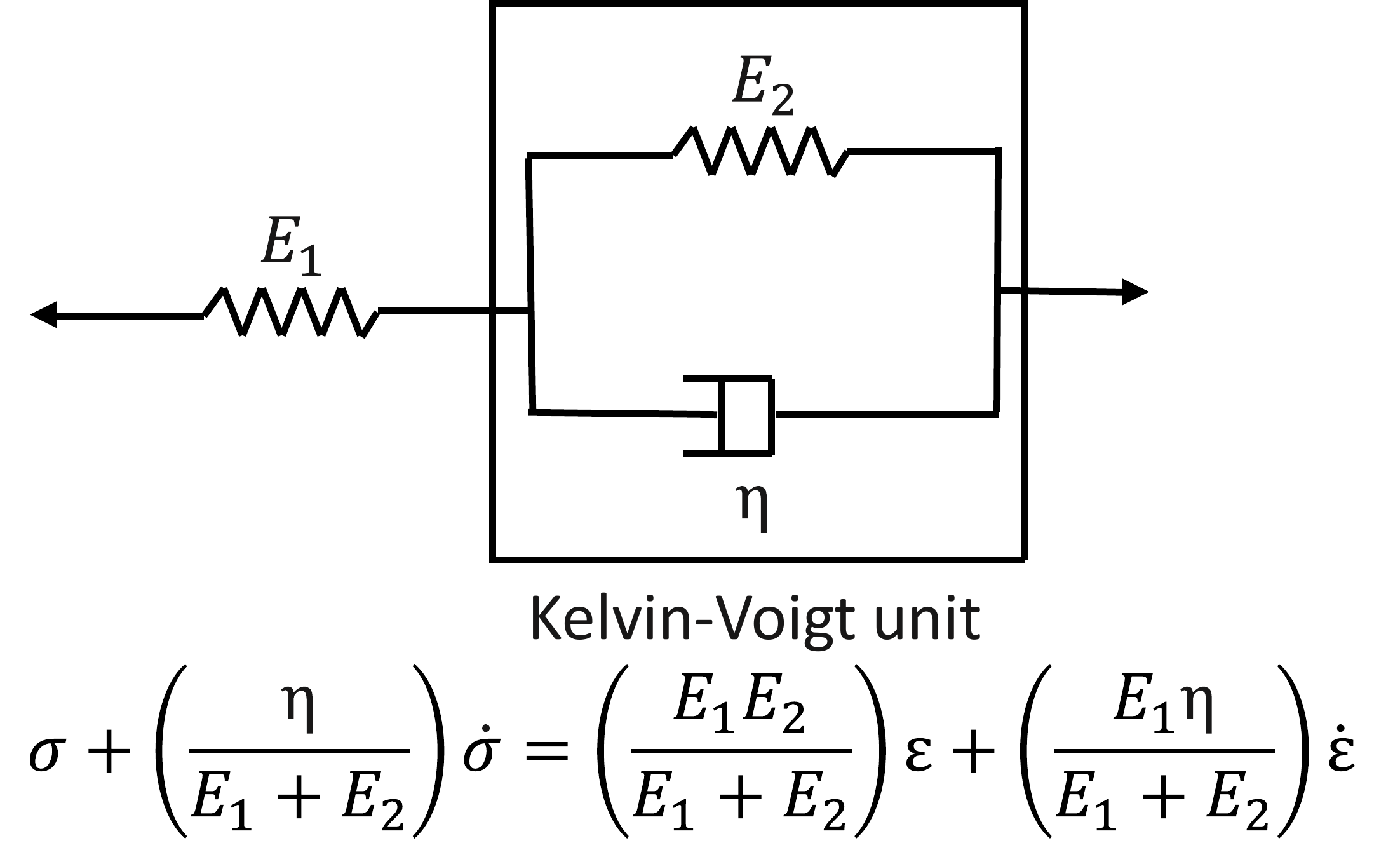}
		\caption{}
		\label{fig:SLS_2}
	\end{subfigure}
	\caption{The rheological networks of a: (a) Maxwell (b) Kelvin-Voigt (c) Zener and, (d) Poynting-Thompson model.}
	\label{fig:rheological cnnection}
\end{figure}

\subsubsection{Maxwell model}\label{sec:finite_maxwell}

To develop our constitutive model, We first need to select an appropriate functional form for the Helmholtz free energy and the rate of dissipation function. Here we confine our attention to only isotropic materials. The Helmholtz free energy is chosen to be a function of the three invariants of the elastic Green strain $\mathbf{E}^e$ pulled back into the reference configuration $\kappa_r(\boldsymbol{\mathcal{B}})$. Typically, the invariants of the right or left Cauchy-Green tensor are used for the Helmholtz free energy, e.g., a Neo-Hookean or a Mooney-Rivlin model. However, since there is a one-to-one correspondence between the invariants of the Green strain tensor and that of the Cauchy-Green tensors~\cite{gurtin2001}, it is reasonable to use the invariants of the Green strain tensor or its elastic and viscous parts to construct the Helmholtz free energy~\cite{anssari2024generalised}. These kinematic variables will be used to construct material models throughout this paper. We use the following invariants of $\mathbf{E}^e$ to construct our Helmholtz free energy: 
\begin{equation}\label{eq:MM_I1_I2}
    I_1=\operatorname{tr}\left(\mathbf{F}^{v^{T}}\,\mathbf{E}^e\,\mathbf{F}^v\right),\quad~ I_2=\operatorname{tr}\left(\mathbf{F}^{v^{T}}\,\mathbf{E}^e\,\mathbf{F}^v\right)^2,\quad \text{and}~ I_3 = \det\left(\mathbf{F}^{v^{T}}\,\mathbf{E}^e\,\mathbf{F}^v\right).
\end{equation}
In terms of these invariants, the Helmholtz free energy can be written as
\begin{equation}\label{eq:Finite_MM_psi}
    \rho_0\,\psi=\alpha_1I_1+\alpha_2I_2 + \alpha_3I_3,\quad \text{where}\quad \alpha_i=\rho_o\partial\psi/\partial I_i,\, i=1,2,3.
\end{equation}
The rate of dissipation function is chosen as a quadratic function of $\mathbf{\dot{E}}^v$ in accordance with the postulates of Singh and Paul~(2025)~\cite{singh2025}. Thus, the rate of dissipation function reads
\begin{equation}\label{eq:Finite_MM_xi}
    \xi={\eta}\,\mathbf{\dot{E}}^v\boldsymbol{:}\mathbf{\dot{E}}^v
\end{equation}
where $\eta$ is the viscosity of the material. Clearly, the rate of dissipation is non-negative and also a convex function in $\mathbf{\dot{E}}^v$. Now substituting the chosen Helmholtz free energy in Eq.~\eqref{eq:Finite_MM_psi} into Eq.~\eqref{eq:green_elastic_solid} and using Eq.~\eqref{eq:Ee_Ev_and*}, we obtain the second Piola-Kirchhoff stress as
\begin{equation}\label{eq:S_maxwell}
    \mathbf{{{S}}}=\mathbf{\tilde{S}}+\alpha_3I_3\left(\mathbf{E}-\mathbf{E}^v\right)^{-1}\quad\text{with}\quad\mathbf{\tilde{S}}=\alpha_1\mathbf{I}+2\alpha_2\,(\mathbf{E}-\mathbf{E}^v).
\end{equation}
Here $\mathbf{\tilde{S}}$ acts as the deviatoric part of the second Piola-Kirchhoff stress. The evolution equation for $\mathbf{E}^v$ is obtained by substituting the rate of dissipation function into \eqref{eq:evoluation_eqn_E_v} as
\begin{equation}\label{eq:evolution_maxwell}
    \mathbf{\dot{E}}^v=\dfrac{1}{\eta}\left(\alpha_1\mathbf{I}+2\alpha_2\,(\mathbf{E}-\mathbf{E}^v)+\alpha_3I_3\left(\mathbf{E}-\mathbf{E}^v\right)^{-1}\right).
\end{equation}
One can easily notice the similarity between the (linearized) evolution equations of Lubliner~(1985)~\cite{lubliner1985model}, Reese and Govindjee~\cite{reese1998theory} and Eq.~\eqref{eq:evolution_maxwell} with some additional terms associated with the material parameters $\alpha_1$ and $\alpha_3$. Now combining Eqs.~\eqref{eq:Ee_Ev_and*},~\eqref{eq:S_maxwell} and~\eqref{eq:evolution_maxwell}, we arrive at the governing equation for a Maxwell material as
\begin{equation}\label{eq:Finite_MM_evoluation_equation}
   \mathbf{\dot{\tilde{S}}}+\left({2\alpha_2}/{\eta}\right)\,\mathbf{S}=2\alpha_2 \mathbf{\dot{E}} .
\end{equation}
It is worth noting that the model proposed by Bonet~(2001)~\cite{bonet2001large,liu2025large} emerges as a special case of Eq.~\eqref{eq:Finite_MM_evoluation_equation} under the condition of a constant deformation gradient, i.e., when $\mathbf{\dot{E}}=0$. We further note that the developed governing equation is slightly different from that developed by Rajagopal and Srinivasa~(2001)~\cite{rajagopal2001modeling} or the traditionally used upper-convected Maxwell model. This is due to the fact that while an objective stress rate appears in these governing equations, our developed model involves a simple material time-derivative of the second Piola-Kirchhoff stress. This is a direct consequence of using a Lagrangian framework. Since the kinetic and kinematic variables used here, such as the second Piola-Kirchhoff stress and the Green strain remain the same under a transformation of the material frame,  their rates are also the same under the same transformation. On the other hand, the Cauchy stress transforms according to $\mathbf{T}^*=\mathbf{Q}\,\mathbf{T}\,\mathbf{Q}^T, \mathbf{Q}\in\boldsymbol{\mathcal{O}}^+(3)$ where $\mathbf{T}^*$ and $\mathbf{T}$ represent the Cauchy stresses, measured by two observers whose frames of reference differ by a rigid body motion. Therefore, to ensure form invariance of the constitutive relationship under material frame transformation, an objective stress rate is needed for an Eulerian frame of reference~\cite{jog2015continuum}. In contrast, a material time derivative of the second Piola-Kirchhoff stress or the Green strain is sufficient to preserve the constitutive form invariance for a Lagrangian formulation. 

For an incompressible material, while the term associated with the third invariant of $\mathbf{E}^e$ needs to be dropped from the expression of $\psi$, two additional constraints of $J=1$ and $J^v=1$ must be incorporated. Therefore, the Helmholtz free energy and the rate of dissipation function take the form
\begin{equation}\label{eq:Finite_MM_psi_Inc}
    \psi=\alpha_1I_1+\alpha_2I_2,\quad \text{and}\quad \xi={\eta}\,\mathbf{\dot{E}}^v\boldsymbol{:}\mathbf{\dot{E}}^v.
\end{equation}
Substituting $\psi$ and $\xi$ from Eq.~\eqref{eq:Finite_MM_psi_Inc} into  \eqref{eq:green_inc} and~\eqref{eq:evoluation_eqn_E_v_InC}, we arrive at
\begin{equation}\label{eq:Finite_MM_evolution_equation_Inc}
   \eta\,\mathbf{\dot{\tilde{S}}}+{2\alpha_2}\,\mathbf{\tilde{S}}=2\alpha_2\,\eta\, \mathbf{\dot{E}}+{2\alpha_2}\,q\,J^v\,\mathbf{C}^{v^{-1}}
   \end{equation}
   \text{with}
   \begin{equation}
       \mathbf{\tilde{S}}=\alpha_1\,\mathbf{I}+2\alpha_2\,(\mathbf{E}-\mathbf{E}^v)\quad\text{and}\quad\mathbf{S}=\mathbf{{\tilde{S}}}-pJ\mathbf{C}^{-1}.
   \end{equation}
The governing equation for an incompressible Maxwell model is similar to that of a compressible material with a notable exception of the last term in Eq.~\eqref{eq:Finite_MM_evolution_equation_Inc}.

\subsubsection{Zener model}\label{Sec:Finite_deformation_SLS_model}

We now extend this theory to derive the constitutive model for a Zener model, shown in Fig.~\ref{fig:SLS_1}. For a compressible Zener model, the Helmholtz free energy is assumed to depend on three additional invariants of the total strain $\mathbf{E}$. This choice of the Helmholtz free energy is based on the idea of decomposing it into its equilibrium (function of $\mathbf{E}$) and non-equilibrium (function of $\mathbf{E}^e$) parts~\cite{reese1998theory,hossain2012experimental,kumar2016two,sadik2024nonlinear}. The three additional invariants are defined as
\begin{equation}\label{eq:SLS_I1_I2}
    \begin{aligned}
        {I}^s_1&=\operatorname{tr}\left(\mathbf{E}\right),\quad {I}^s_2=\operatorname{tr}\left(\mathbf{E}\right)^2,\quad \text{and}\quad {I}^s_3 = \operatorname{det}\left(\mathbf{E}\right).
    \end{aligned}
\end{equation}
With these invariants, the Helmholtz free energy is written as
\begin{equation}\label{eq:Finite_SS_psi}
        \rho_0\,\psi=\alpha_1\,I_1+\alpha_2\,I_2+\alpha_3\,I_3+\alpha_4\,{I}^s_1+\alpha_5\,I^s_2+\alpha_6\,I^s_3
\end{equation}
where $\alpha_i=\rho_0\,\partial\psi/\partial I_i$ and $\alpha_{i+3}=\rho_o\,\partial\psi/\partial I^s_i$, $i=1,2,3$. The rate of dissipation function is chosen to have the same form as that of a Maxwell model. Substituting Eq.~\eqref{eq:Finite_SS_psi} and~\eqref{eq:Finite_MM_xi} into Eq.~\eqref{eq:green_elastic_solid} and \eqref{eq:evoluation_eqn_E_v}, we obtain
\begin{equation}\label{eq:Finite_SLS_evoluation_equation}
\eta\,\mathbf{\dot{\tilde{S}}}+{2\alpha_2}\,\mathbf{S}={2\alpha_2\,\alpha_4}\,\mathbf{I}+2(\alpha_2+\alpha_5)\,\eta\,\mathbf{\dot{E}}+{4\alpha_2\,\alpha_5}\,\mathbf{E}+{2\alpha_2\,\alpha_6}\,\operatorname{det}\left(\mathbf{E}\right)\,\mathbf{E}^{-1}
\end{equation}
where
\begin{equation}
    \mathbf{\tilde{S}}=\mathbf{S}-\alpha_3\,I_3\,(\mathbf{E}-\mathbf{E}^v)^{-1}-\alpha_6\,I^s_3\,\mathbf{E}^{-1}\quad \text{with}\quad \mathbf{\tilde{S}}=(\alpha_1+\alpha_4)\,\mathbf{I}+2(\alpha_2+\alpha_5)\,\mathbf{E}-2\alpha_2\,\mathbf{E}^v.
\end{equation}

For an incompressible Zener solid, we follow the same strategy as for an incompressible Maxwell model. The third invariants of $\mathbf{E}$ and $\mathbf{F}^{v^T}\,\mathbf{E}^e\,\mathbf{F}^v$ are dropped from the expression of the Helmholtz free energy which leads to
\begin{equation}\label{eq:Finite_SS1_psi_Inc}
       \rho_0\, \psi=\alpha_1\,I_1+\alpha_2\,I_2+\alpha_4\,{I}^s_1+\alpha_5\,I^s_2.
\end{equation}
Similar to the compressible material, here we take the same expression for the rate of dissipation function as in the case of an incompressible Maxwell material. The constraints for incompressibility $J=1$ and $J^v=1$ must be incorporated in the evolution equation and the expression of the second Piola-Kirchhoff stress. This is achieved by using the appropriate equations~\eqref{eq:evoluation_eqn_E_v_InC} and ~\eqref{eq:green_inc}, derived for the incompressible materials. Now combining Eqs.~\eqref{eq:Ee_Ev_and*}, \eqref{eq:evoluation_eqn_E_v_InC}, \eqref{eq:green_inc} and \eqref{eq:Finite_SS1_psi_Inc}, we obtain
\begin{equation}\label{eq:Finite_SLS_evoluation_equation_Inc}
\eta\,\mathbf{\dot{\tilde{S}}}+{2}\alpha_2\,\mathbf{\tilde{S}}={2}\alpha_2\alpha_4\,\mathbf{I}+2\eta\,(\alpha_2+\alpha_5)\,\mathbf{\dot{E}}+{4}\alpha_2\,\alpha_5\,\mathbf{E}+{2}\,\alpha_2\,q\,J^v\,\mathbf{C}^{v^{-1}}.
\end{equation}
Here the expressions for the second Piola-Kirchhoff stress and its deviatoric part read
\begin{equation}
\mathbf{S}=\mathbf{\tilde{S}}-p\,J\,\mathbf{C}^{-1}\quad\text{with}\quad\mathbf{\tilde{S}}=(\alpha_1+\alpha_4)\,\mathbf{I}+2(\alpha_2+\alpha_5)\,\mathbf{E}-2\alpha_2\,\mathbf{E}^v.
\end{equation}

\section{A stress space formulation of the evolving natural configurations framework}\label{sec:Configurational_force}
We develop the stress-space formulation of the theory of natural configurations following the works of Rajagopal and Srinivasa~(1998,2004)~\cite{rajagopal1998mechanics, rajagopal2000thermodynamic} in this section. This formulation is essential for our subsequent development of the nonlinear Kelvin-Voigt type models.
\subsection{Configurational forces}
For a stress space formulation, we start with calculating the force on an inhomogeneity, known as a configurational or a driving force, following Eshelby's procedure~\cite{eshelby1957determination} which was recast into the evolving natural configurations framework by Rajagopal and Srinivasa ~(2005)~\cite{rajagopal2005role}. In this method, the configurational force is determined by observing the change in the total strain energy in a body due to a small perturbation in the natural configuration. Since the Helmholtz free energy $\psi$ represents the stored energy per unit volume of the undeformed configuration of the body, the total strain energy is given by
\begin{equation}
    {\Psi}=\int_{\Omega}{}\rho_0\,\overline{\psi} (\mathbf{E}, \mathbf{E}^v)\, dV.
\end{equation}
Now let us provide a small perturbation in the natural configuration (i.e., $\mathbf{E}^v$) with any tensor-valued function $\alpha\,\mathbf{H}$, where $\alpha\in \mathbb{R}$ and $\alpha\ll1$. Thus, the strain energy per unit volume after perturbation is 
\begin{equation}\label{eq:psi_deformation}  \psi=\bar{\psi}\left(\mathbf{E},\mathbf{E}^v+\alpha\mathbf{H}\right).
\end{equation}
The change in the total strain energy due to the perturbation, after the terms of the order $\mathcal{O}(\alpha^2)$ and higher are neglected, reads
\begin{equation}\label{eq:configurational_psi}
    -\left.\dfrac{\partial\Psi}{\partial\alpha}\right|_{\alpha=0}=-\int_{\Omega} \left( \rho_0\frac{\partial \bar{\psi}}{\partial \mathbf{E}^v} \boldsymbol{:} \mathbf{H}\right) \, dV.
\end{equation}
Since $\alpha\,\mathbf{H}$ signifies the perturbation in the natural configuration in Eq.~\eqref{eq:configurational_psi}, following Eshelby's~(1957)~\cite{eshelby1957determination} definition, the driving (configurational) force behind the changes in the natural configuration is given by
\begin{equation}\label{eq:viscous_configurational_force}
    \quad \mathbf{A}^v\defeq-\rho_0\dfrac{\partial\bar{\psi}}{\partial{\mathbf{E}}^v}.
\end{equation}
This configurational force can be shown to be equivalent to the Eshelby's energy-momentum tensor~\cite{rajagopal2005role,paul2022use}. 

It is important to note here that the definition of the configurational force depends on the chosen functional form for the Helmholtz free energy, $\psi$. For example, Rajagopal and Srinivasa~(1998)~\cite{rajagopal1998mechanics} chose the Helmholtz free energy of the form $\psi=\tilde{\psi} (\mathbf{F}, \mathbf{F}^v)$ which led to the expression for the configurational force as $\mathbf{A}^v=-\rho_0\,({\partial\tilde{\psi}}/{\partial\mathbf{F}^v})\,\mathbf{F}^{v^T}$. In their derivation of a Kelvin-Voigt type standard solid model, Huber and Tsakmakis~(2000)~\cite{huber2000finite} used the form of $\psi$ to be $\psi=\hat{\psi}(\mathbf{F}^e, \mathbf{F}^v)$. In this case, the expression for the configurational force can be determined by the same exercise with some important modifications. Let us choose a small perturbation in $\mathbf{F}^e$ and $\mathbf{F}^v$ \emph{both} in such a way that the total deformation gradient remains unaltered. With these perturbations, the Helmholtz free energy reads 
\begin{equation}
    \psi=\hat{\psi}(\mathbf{F}^e-\alpha\,\mathbf{F}^e\,\mathbf{H},\mathbf{F}^v+\alpha\,\mathbf{H}\,\mathbf{F}^v)\quad\text{with}\quad \alpha\ll1.
\end{equation}
One can easily verify that a multiplication of the perturbed elastic and viscous deformation gradients results in the total deformation gradient $\mathbf{F}$ if the term of the order $\mathcal{O}(\alpha^2)$ is neglected. Now the configurational force is obtained from the variation of the total strain energy as
\begin{equation}
-\dfrac{\partial\hat{\Psi}}{\partial \alpha}\bigg|_{\alpha=0}=\int_{\Omega}\underbrace{\rho_0\left(\mathbf{F}^{e^T}\,\dfrac{\partial\hat{\psi}}{\partial \mathbf{F}^e}-\dfrac{\partial \hat{\psi}}{\partial \mathbf{F}^v}\,\mathbf{F}^{v^T}\right)}_{\mathbf{A}^v}\boldsymbol{:}\,\mathbf{H}\:dV.
\end{equation}

\subsection{Constitutive relations}

Using the configurational forces derived above, we now formulate the required constitutive relations within the stress space formulation. For a compressible material, the constraint equation~\eqref{eq:xi_constraint} for the rate of dissipation can be alternatively written as 
\begin{equation}\label{eq:constrained_rajagopal}
\xi=\mathbf{A}^v\boldsymbol{:}\mathbf{\dot{{E}}}^v.
\end{equation}
Now following the exercise of a constrained maximization of the rate of dissipation function, similar to the one shown in \S~\ref{sec:constitutive_relation}, we obtain
\begin{equation}\label{eq:Ev_xi_evoluation}
    \dfrac{\partial\xi}{\partial\mathbf{A}^v}=\overline{\lambda}\,\mathbf{\dot{{E}}}^v
\end{equation}
where $\overline{\lambda}$ is the Lagrange multiplier that can be determined by the satisfaction of the constraint equation~\eqref{eq:constrained_rajagopal} as
\begin{equation}\label{eq:Lagrange_stress_space}
    \overline{\lambda}=\dfrac{\mathbf{A}^v\boldsymbol{:}\partial\xi/\partial\mathbf{A}^v}{\xi}
\end{equation}

For an incompressible material, the constraints $J=1$ and $J^v=1$ must be incorporated through the modification of the Helmholtz free energy as $\psi_{inc}=\psi-p\,(J-1)-q\,(J^v-1)$. This leads to the modified definition of the configurational force as
\begin{equation}\label{eq:configurational_force_inc}
    \mathbf{A}^v_{inc}=-\rho_0\,\dfrac{\partial \psi}{\partial \mathbf{E}^v}+q\,J^v\,\mathbf{C}^{v^{-1}}.
\end{equation}
Now following the same procedure as before, it can be shown that the evolution equation for incompressible material remains the same while the difference due to the constraints of incompatibility is only apparent in the definition of the configurational force.

\subsection{Modeling of Kelvin-Voigt type materials}\label{sec:applications_stress_space}

We now use the stress-space formulation to develop constitutive models for Kelvin-Voigt type solids. The significant difference between the Maxwell-type and Kelvin-Voigt-type models arises from the choice of the rate of dissipation function. First we develop a Kelvin-Voigt type standard solid model, also known as a Poynting-Thompson model. After that, a Kelvin-Voigt model is obtained as a limiting case of the standard solid models.

\subsubsection{Poynting-Thompson model}\label{Sec:Finite_deformation_SLS_KV_model}
Let us now start with the derivation of a finite deformation version of the Poynting-Thompson model shown in Fig.~\ref{fig:SLS_2}. Following the idea of Huber and Tsakmakis~(2000)~\cite{huber2000finite}, the Helmholtz free energy for this model is chosen as
\begin{equation}\label{eq:Finite_SLS_psi}
    \rho_0\,\psi=\alpha_1\,I_1+\alpha_2\,I_2+\alpha_3\,I_3+\alpha_4\,I^k_1+\alpha_5\,I^k_2+\alpha_6\,I^k_3
\end{equation}
where $I_1, I_2, I_3$ are the invariants defined in Eq.~\eqref{eq:MM_I1_I2}. The additional invariants are given as
\begin{equation}\label{eq:KV_I1_I2}
I^k_1=\operatorname{tr}\left(\mathbf{E}^v\right),\quad I^k_2=\operatorname{tr}\left(\mathbf{E}^{v^{2}}\right),\quad\text{and}~I^k_3=\operatorname{det}\left(\mathbf{E}^v\right).
\end{equation}
Here $\alpha_i=\partial\psi/\partial I_i$ and $\alpha_{i+3}=\partial\psi/\partial I^k_i, i=1,2,3$. From the chosen form for the Helmholtz free energy, we can obtain the second Piola-Kirchhoff stress and the configurational force from Eqs.~\eqref{eq:green_elastic_solid} and~\eqref{eq:viscous_configurational_force}, respectively as
\begin{equation}\label{eq:finite_SLS_stress}
     \mathbf{S}=\rho_0\,\dfrac{\partial\psi}{\partial\mathbf{E}}=\alpha_1\mathbf{I}+2\alpha_2\left(\mathbf{E}-\mathbf{E}^v\right)+ \alpha_3I_3\left(\mathbf{E}-\mathbf{E}^v\right)^{-1} \implies\tilde{\mathbf{S}}=\alpha_1\mathbf{I}+2\alpha_2\left(\mathbf{E}-\mathbf{E}^v\right)
\end{equation}
\text{and,}
\begin{equation}\label{eq:finite_SLS__viscous_force}
\mathbf{A}^v=-\rho_0\,\dfrac{\partial\psi}{\partial\mathbf{E}^v}=\mathbf{S}-\alpha_4\mathbf{I}-2\alpha_5\,\mathbf{E}^v-\alpha_6\,I^k_3\,\mathbf{E}^{v^{-1}}
\end{equation}
where the deviatoric stress $\tilde{\mathbf{S}}$ is defined as $\tilde{\mathbf{S}}\defeq\mathbf{S}-\alpha_3I_3\left(\mathbf{E}-\mathbf{E}^v\right)^{-1}$. Since $\mathbf{A}^v$ is the driving force for the evolution of the natural configuration, we choose the rate of dissipation function in accordance with Singh and Paul~(2025)~\cite{singh2025} as
\begin{equation}\label{eq:finite_SLS_xi}
    \xi=\dfrac{1}{\eta}\,\mathbf{A}^v\boldsymbol{:}\mathbf{A}^v.
\end{equation}
Since the chosen rate of dissipation is a quadratic function of the configurational force $\mathbf{A}^v$ with the viscosity parameter $\eta>0$, it can be easily verified that the rate of dissipation function $\xi$ is convex in the configurational force space. Moreover, it satisfies the requirements stated in Theorem 3 of Rajagopal and Srinivasa~(1998)~\cite{rajagopal1998inelastic}. The choice of the functional form of the rate of dissipation function is a significant departure from the existing viscoelastic models. Now using Eqs.~\eqref{eq:Ev_xi_evoluation} and~\eqref{eq:finite_SLS_xi} on the chosen rate of dissipation function, one can obtain the evolution equation for the viscous strain. From Eq.~\eqref{eq:Lagrange_stress_space}, the Lagrange multiplier is evaluated as $\overline{\lambda}=2$. Therefore, substituting the value of $\overline{\lambda}$ and the expression for the driving force $\mathbf{A}^v$ from Eq.~\eqref{eq:finite_SLS__viscous_force} into Eq.~\eqref{eq:Ev_xi_evoluation}, the evolution equation for $\mathbf{E}^v$ is obtained as
\begin{equation}\label{eq:KV_SLS_evolution}
    \mathbf{\dot{E}}^v=\dfrac{1}{\eta}\left(\mathbf{S}-\alpha_4\mathbf{I}-2\alpha_5\,\mathbf{E}^v-\alpha_6\,I^k_3\,\mathbf{E}^{v^{-1}}\right).
\end{equation}
Now taking a material time derivative of Eq.~\eqref{eq:finite_SLS_stress} and using Eqs.~\eqref{eq:Ee_Ev_and*} and \eqref{eq:KV_SLS_evolution}, we obtain the governing equation for a Poynting-Thompson solid as
\begin{equation}\label{eq:Finite_SLS_governing_equation}
    \eta\,\mathbf{\dot{\tilde{S}}}+{2\,\alpha_2}\,\mathbf{{S}}=2\,\alpha_2\,\eta\,\mathbf{\dot{E}}+4\alpha_2\,\alpha_5\,\mathbf{E}^v+{2\,\alpha_2\,\alpha_4}\,\mathbf{I}+{2}\alpha_2\,\alpha_6\,\mathbf{I}^k_3\,\mathbf{E}^{v{-1}}.
\end{equation}


For an incompressible material, the terms associated with $I_3$ and $I^k_3$ are dropped and thus, the expression for the Helmholtz free energy reads
\begin{equation}\label{eq:Finite_SS2_psi_Inc}
        \rho_0\,\psi=\alpha_1\,I_1+\alpha_2\,I_2+\alpha_4\,{I}^k_1+\alpha_5\,I^k_2.
\end{equation}
The second Piola-Kirchhoff stress and the configurational force can be obtained from this Helmholtz free energy using Eqs.~\eqref{eq:green_inc} and~\eqref{eq:configurational_force_inc} respectively, as
\begin{equation}\label{eq:PK2_SLS_KV}
    \mathbf{{{S}}}=\mathbf{\tilde{S}}-p\,J\,\mathbf{C}^{-1}\quad\text{with}\quad\mathbf{\tilde{S}}=\alpha_1\mathbf{I}+2\alpha_2\,(\mathbf{E}-\mathbf{E}^v)
\end{equation}   
\text{and,}
\begin{equation}\label{eq:configurational_force_SLS_KV}
    \mathbf{A}^v=\mathbf{\tilde{S}}-\alpha_4\mathbf{I}-2\alpha_5\,\mathbf{E}^v+q\,J^v\,\mathbf{C}^{v^{-1}}.
\end{equation}
Now choosing the same expression for the rate of dissipation function as in Eq.~\eqref{eq:finite_SLS_xi} along with the modified definition of the configurational force in Eq.~\eqref{eq:configurational_force_SLS_KV} and using Eqs.~\eqref{eq:PK2_SLS_KV}, \eqref{eq:Ee_Ev_and*} and \eqref{eq:Ev_xi_evoluation}, we obtain
\begin{equation}\label{eq:Finite_SS2_governing_equation_Inc}
    \eta\,\mathbf{\dot{\tilde{S}}}+{2\,\alpha_2}\mathbf{\tilde{S}}=2\,\alpha_2\,\eta\,\mathbf{\dot{E}}+{4}\alpha_2\,\alpha_5\,\mathbf{E}^v+{2}\alpha_2\,\alpha_4\,\mathbf{I}- {2}\alpha_2\,q\,J^v\,\mathbf{C}^{v^{-1}}.
\end{equation}
Eq.~\eqref{eq:Finite_SS2_governing_equation_Inc} is the required governing equation for an incompressible Poynting-Thompson model. It is important to note here that the governing equations~\eqref{eq:Finite_SLS_evoluation_equation} and~\eqref{eq:Finite_SLS_evoluation_equation_Inc} for a Zener model is significantly different from the ones for a Poynting-Thompson solid, given in Eqs.~\eqref{eq:Finite_SLS_governing_equation} and~\eqref{eq:Finite_SS2_governing_equation_Inc}, respectively. This difference has a significant consequence which will be discussed next.

\subsubsection{Limiting case: Kelvin-Voigt model}\label{sec:limiting}

From the Poynting-Thompson model, one can easily observe that for a limiting case of $\alpha_4, \alpha_5, \alpha_6\rightarrow 0$, the governing equation~\eqref{eq:Finite_SLS_governing_equation} reduces to that of a Maxwell model, i.e., Eq.~\eqref{eq:Finite_MM_evoluation_equation}. This result is due to the fact that the limits under consideration render the assumed form for $\psi$ as well as the resulting configurational force of a Poynting-Thompson model same as that of a Maxwell model. This, in turn, results in the same evolution equations for both the models even though the rate of dissipation functions are chosen to be different. Motivated by this observation, we now examine the limiting case for a Zener model.

Let us consider the material parameters $\alpha_1$, $\alpha_2$ and $\alpha_3$ for the Zener model to be much higher than the other material parameters. For a small-strain rheological model (Fig.~\ref{fig:SLS_1}), this limit is tantamount to having the spring with modulus $E_2$ too stiff such that it acts like a rigid element. Therefore, under this condition, the Zener model (Fig.~\ref{fig:SLS_1}) shows an equivalent response to that of a Kelvin-Voigt model (Fig.~\ref{fig:KV}). Now let us explore the implication of this limit for the finite deformation Zener model derived in \S~\ref{Sec:Maxwell_type_material}.\ref{Sec:Finite_deformation_SLS_model}. Dividing both sides of the governing equation~\eqref{eq:Finite_SLS_evoluation_equation} of the Zener model by $2\alpha_2$ and considering $\alpha_2$ to be very large, the governing equation reduces to
\begin{equation}\label{eq:KV_compressible_governing}
\mathbf{S}=\alpha_4\,\mathbf{I}+2\alpha_5\,\mathbf{E}+\alpha_6\,I^s_3\,\mathbf{E}^{-1}+\eta\,\mathbf{\dot{E}}.
\end{equation}
Eq.~\eqref{eq:KV_compressible_governing} is the desired finite deformation version of the Kelvin-Voigt model in its Lagrangian form. In his model, Rajagopal~(2009)~\cite{rajagopal2009note} considered the Kelvin-Voigt material as a mixture between a neo-Hookean elastic solid and a Newtonian fluid which led to the constitutive assumption of
\begin{equation}\label{eq:KV_Rajagopal}
    \mathbf{T}=\mu\,\mathbf{B}-p\,\mathbf{I}+\eta\,\mathbf{D}.
\end{equation}
One can easily notice the similarity between~\eqref{eq:KV_Rajagopal} and our model with one important distinction. From the governing equation~\eqref{eq:KV_compressible_governing}, our Kelvin-Voigt model can be interpreted as a mixture between a variant of a \emph{Mooney-Rivlin elastic} solid and a Newtonian fluid. It is worth noting that our Kelvin-Voigt model is obtained as a direct consequence of the theory of evolving natural configurations and its interpretation as a mixture is not necessary for its derivation. We conclude this section by providing a Kelvin-Voigt model for an incompressible material. In view of the above interpretation, $J=1$ can be considered as the only constraint of incompressibility here. Thus, from Eq.~\eqref{eq:Finite_SLS_evoluation_equation_Inc}, we obtain the required governing equation as
\begin{equation}\label{eq:KV_incompressible_governing}
\mathbf{S}=\alpha_4\,\mathbf{I}+2\alpha_5\,\mathbf{E}-p\,J\,\mathbf{C}^{-1}+\eta\,\mathbf{\dot{E}}.
\end{equation}
While the constitutive equations for the other viscoelastic models are implicit, the constitutive equations for the Kelvin-Voigt model can be directly integrated since they are explicit equations in the total deformation gradient, $\mathbf{F}$, or the resulting Green strain and its rate.

\section{Integration algorithms and numerical examples}\label{sec:numerical_example}

To show the efficacy of our developed models, we develop integration algorithms and demonstrate the utility of the models through some simple boundary value problems in this section. Although development of a full-scale finite element framework is required to exploit the full potential of the developed models, such framework is beyond the scope of the current work and will be addressed in the future.

\subsection{Numerical implementation of the developed models}\label{sec:Numerical_implementation}

As mentioned earlier, the governing equations for the Maxwell, Zener and Poynting-Thomposon models are implicit equations, whereas the governing equation for the Kelvin-Voigt model is explicit in the Green strain $\mathbf{E}$ and its material time-derivative. While the governing equation for the Kelvin-Voigt model can be directly integrated, a suitable numerical method must be adopted for the other three. Let us first explore the steps involved in the numerical solutions for these models. Here we assume that the test is displacement-controlled and thus, the history of the total deformation gradient $\mathbf{F}$ is known at all time. The integration algorithms for these models proceed as follows:


\emph{1. Discretization of the total time interval:} Let us start with the discretization of the total time of the test $[0,T]$ into smaller intervals such that $\bigcup_{n=1}^k[t_n,t_{n+1}]=[0,T]$. Here $h$ is a small time increment or a time step such that $t_{n+1}=t_n + h$ and $k$ denotes the total number of time intervals. At the previous time step $t_n$, the deformation gradient $\mathbf{F}_n$ as well as all other variables such as $\mathbf{S}_n$, $\mathbf{E}^v_n$ etc. are known. Since the test is assumed to be displacement-controlled, the increment in the deformation gradient and thus, its value at the current time step, $\mathbf{F}_{n+1}$ is also known. The total Green strain $\mathbf{E}_{n+1}$ and its rate $\mathbf{\dot{E}}_{n+1}$ can be evaluated from the known values of the deformation gradient, using Eq.~\eqref{eq:Green-Lagrangian strain} and through the equation 
\begin{equation}\label{eq:Green_strain_rate_numerical}
    \mathbf{\dot{E}_{n+1}}=\dfrac{\mathbf{E}_{n+1}-\mathbf{E}_n}{h}+\mathcal{O}(h^2).
\end{equation}

\emph{2. Numerical integration of the evolution equations:} With these information, now we proceed to a numerical integration of the evolution equations~\eqref{eq:evolution_maxwell} and~\eqref{eq:KV_SLS_evolution} for the Maxwell/Zener and Poynting-Thompson model respectively. Here we use a Heun's method which is more amenable for our purpose. This method uses an explicit forward Euler predictor with a trapezoidal corrector and is accurate up to the second order. With the known values of $\mathbf{E}_n$ and $\mathbf{E}^v_n$, we first predict the value of $\mathbf{E}^v_{n+1}$ as
\begin{equation}
    \mathbf{\bar{E}}^v_{n+1} = \mathbf{E}^v_n + h f\left(\,\mathbf{E}_n,\,\mathbf{E}^v_n\right).
\end{equation}
The function $f(\mathbf{E},\mathbf{E}^v)$ for the Maxwell/Zener model is given as
\begin{equation}
    f\left(\,\mathbf{E},\,\mathbf{E}^v\right) = \dfrac{1}{\eta}\left[\alpha_1\mathbf{I}+2\alpha_2\left(\mathbf{E}-\mathbf{E}^v\right)+ \alpha_3\operatorname{det}\left(\mathbf{E}-\mathbf{E}^v\right)\left(\mathbf{E}-\mathbf{E}^v\right)^{-1}\right]
\end{equation}
whereas for the Poynting-Thompson model, it reads
\begin{equation}
\begin{aligned}
f\left(\,\mathbf{E},\,\mathbf{E}^v\right) = \dfrac{1}{\eta}[\alpha_1\mathbf{I}+2\alpha_2\left(\mathbf{E}-\mathbf{E}^v\right)+ \alpha_3\operatorname{det}\left(\mathbf{E}-\mathbf{E}^v\right)\left(\mathbf{E}-\mathbf{E}^v\right)^{-1}-\alpha_4\mathbf{I}\\-2\alpha_5\,\mathbf{E}^v-\alpha_6\,\operatorname{det}\left(\mathbf{E}^v\right)\,\mathbf{E}^{v^{-1}}].
\end{aligned}
\end{equation}
$\mathbf{E}^v_{n+1}$ is finally evaluated using the trapezoidal corrector via
\begin{equation}
    \mathbf{E}^v_{n+1} = \mathbf{E}^v_n + \dfrac{h}{2}\left[ f\left(\,\mathbf{E}_n,\,\mathbf{E}^v_n\right)+f\left(\,\mathbf{E}_{n+1},\,\mathbf{\bar{E}}^v_{n+1}\right)\right].
\end{equation}

\emph{3. Calculation of the stresses:} With the kinematic variables evaluated at the current time step $t_{n+1}$, the second Piola-Kirchhoff stress can now be easily calculated at this time step. For the Maxwell and Poynting-Thompson model, the second Piola-Kirchhoff stress is computed as
\begin{equation}
    \mathbf{S}_{n+1}=\alpha_1\mathbf{I}+2\,\alpha_2\left(\mathbf{E}_{n+1}-\mathbf{E}^v_{n+1}\right)+ \alpha_3\operatorname{det}\left(\mathbf{E}_{n+1}-\mathbf{E}^v_{n+1}\right)\left(\mathbf{E}_{n+1}-\mathbf{E}^v_{n+1}\right)^{-1}
\end{equation}
whereas for the Zener model, the expression for the second Piola-Kirchhoff stress becomes  
\begin{equation}
\begin{aligned}
\mathbf{S}_{n+1}=\left(\alpha_1+\alpha_4\right)\mathbf{I}+2\,\alpha_2\left(\mathbf{E}_{n+1}-\mathbf{E}^v_{n+1}\right)+ \alpha_3\operatorname{det}\left(\mathbf{E}_{n+1}-\mathbf{E}^v_{n+1}\right)\left(\mathbf{E}_{n+1}-\mathbf{E}^v_{n+1}\right)^{-1}\\+2\,\alpha_5\,\mathbf{E}_{n+1}+ \alpha_6\operatorname{det}\left(\mathbf{E}_{n+1}\right)\,\mathbf{E}_{n+1}-\mathbf{E}^{v^{-1}}_{n+1}.
\end{aligned}
\end{equation}
For a Kelvin-Voigt model, the second Piola Kirchhoff can be directly computed using steps 1 and 3. The second Piola–Kirchhoff stress can be computed as
\begin{equation}
   \mathbf{S}_{n+1}= \alpha_4\,\mathbf{I} + 2\,\alpha_5\mathbf{E}_{n+1} + \alpha_6\,\operatorname{det}\left(\mathbf{E}_{n+1}\right)\,\mathbf{E}^{-1}_{n+1} + \eta\,\dfrac{\left(\mathbf{E}_{n+1}-\mathbf{E}_n\right)}{h}.
\end{equation}

\subsection{Application to a uniaxial stretch}
We now consider a simple homogeneous deformation such as a uniaxial, isochoric stretch of a solid cylinder to demonstrate the response of our model as well as for experimental verification. For this purpose, let us write the deformation map in a cylindrical polar coordinate system. The motion of a material particle is given by
\begin{equation}
    r = R, \quad \theta = \Theta, \quad z = \Lambda(t)\,Z
\end{equation}
where $ \Lambda(t) $ is a time-dependent stretch function. Since the motion is assumed to be isochoric, the deformation gradient $\mathbf{F}$ is calculated as
\begin{equation}
\mathbf{F}(t) = \text{diag}(1/\sqrt{\Lambda(t)},\ 1/\sqrt{\Lambda(t)},\ \Lambda(t))
\end{equation}
For this motion, the right Cauchy-Green tensor $\mathbf{C}$ and the Green-Lagrange strain tensor $\mathbf{E}$ are obtained as 
\begin{equation}
    \mathbf{C} = \text{diag}\left(\dfrac{1}{\Lambda},\dfrac{1}{\Lambda}, \Lambda^2\right),
\quad  \mathbf{E} = \text{diag}\left(\dfrac{1-\Lambda}{2\Lambda}, \dfrac{1-\Lambda}{2\Lambda}, \dfrac{\Lambda^2 - 1}{2}\right). 
\end{equation}
Now that the history of deformation is known, we can use the numerical algorithms developed in \S~\ref{sec:Numerical_implementation} to illustrate the response of the developed models for different functional forms of history of stretch, $\Lambda(t)$.
\begin{table}[h!]
	\centering
	\caption{Material parameters used in the numerical examples}
	\renewcommand{\arraystretch}{1.2}
	\begin{tabular}{|p{1.5cm}| p{1cm}|p{1cm}|p{1cm}|p{1cm}|p{1cm}|p{1cm}|p{1cm}|}
		\hline
		\text{Material} &  $\alpha_1$   &  $\alpha_2$   & $\alpha_3$   & $\alpha_4$   & $\alpha_5$   & $\alpha_5$ & $\eta$ \vspace{4 pt} \\
		\hline
		A
		& 25   &  15   & 10   & 60   & 40   & 40   & 5000 \\
		\hline
		B
		& 21   &  20   & 20   & 28   & 4   & 5   & 140\\
		\hline
	\end{tabular}
	\label{tab:material_parameter}
\end{table}

\begin{figure}[h!]
	\centering
	\begin{subfigure}{0.48\linewidth}
		\includegraphics[width=\linewidth]{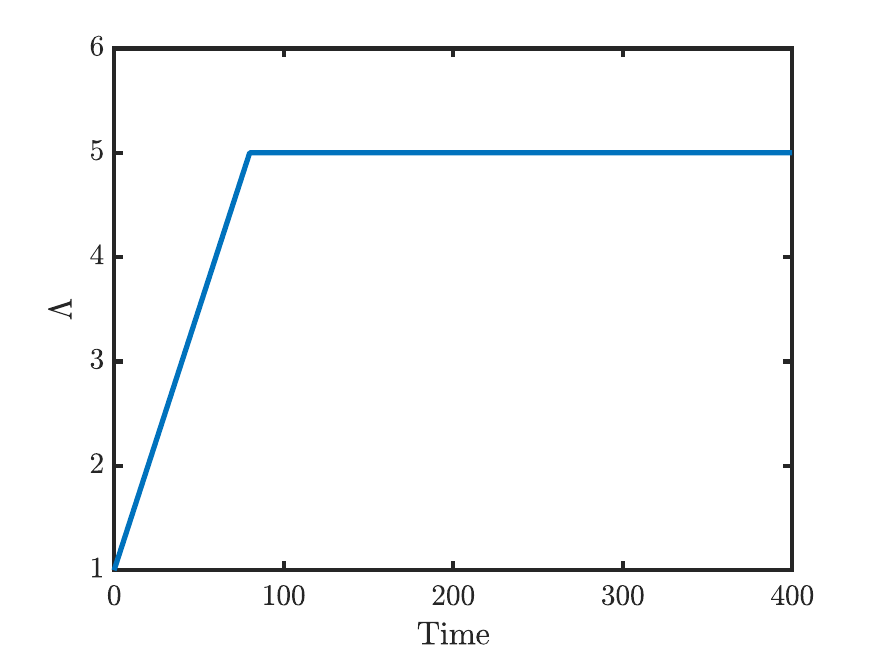}
		\caption{}
		\label{fig:stretch_rate}
	\end{subfigure}
    \begin{subfigure}{0.48\linewidth}
		\includegraphics[width=\linewidth]{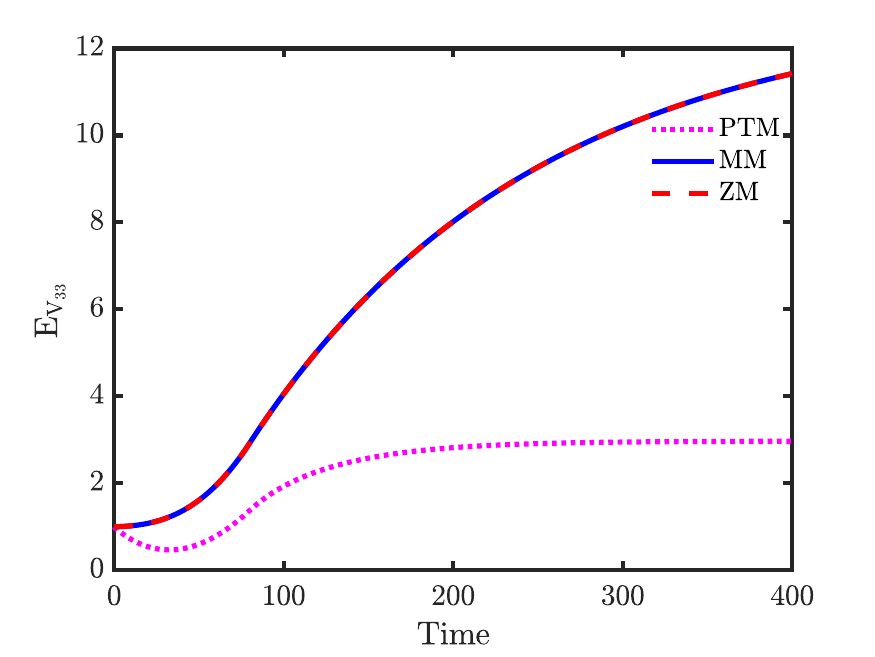}
		\caption{}
		\label{fig:Viscous_strain}
	\end{subfigure}
	\begin{subfigure}{0.48\linewidth}
		\includegraphics[width=\linewidth]{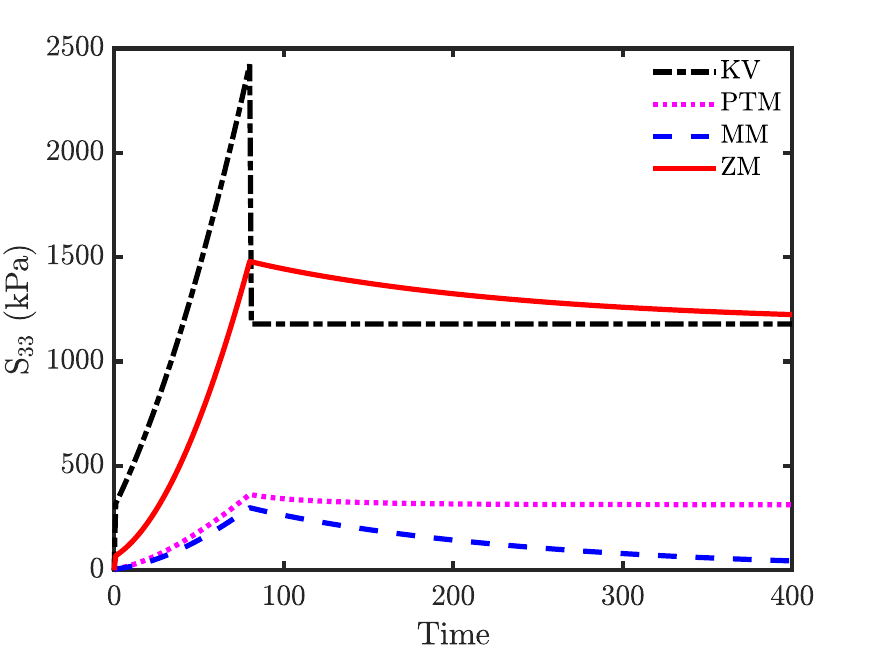}
		\caption{}
		\label{fig:combined_model}
	\end{subfigure}
    \begin{subfigure}{0.48\linewidth}
		\includegraphics[width=\linewidth]{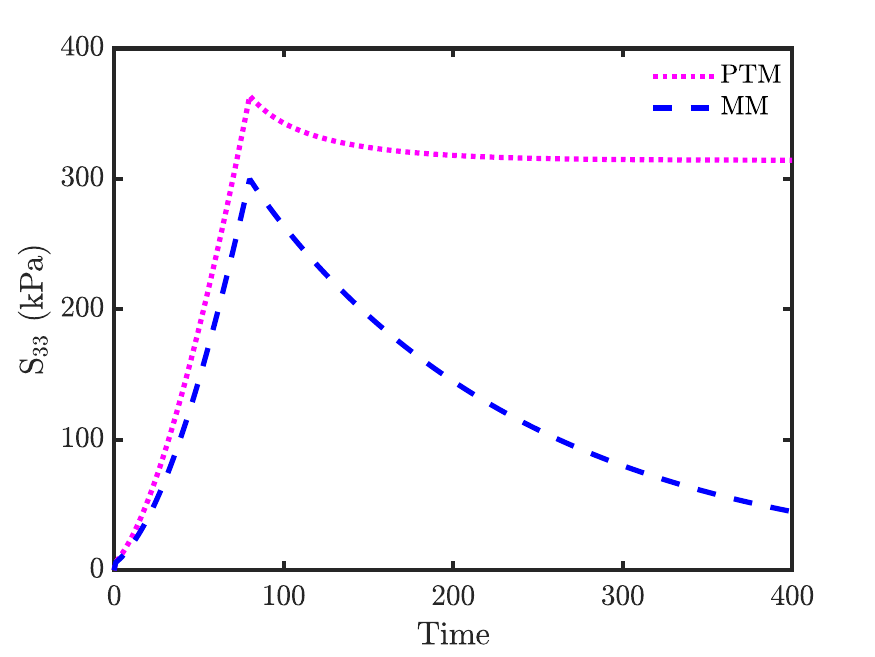}
		\caption{}
		\label{fig:PTM_MM}
	\end{subfigure}
	\caption{(a) Applied stretch with respect to time, (b) evolution of the viscous strain for the Maxwell, Zener and Poynting-Thompson models, (c) stress response and relaxation behavior of the developed models, (d) enlarged version of the stress response of the Poynting-Thompson and the Maxwell models.}
	\label{fig:rheological cnnection}
\end{figure}

\textbf{\textbullet Response of the developed models:} To investigate the stress relaxation behavior and the evolution of the viscous strain, $\mathbf{E}^v$ of the developed models, we apply a linear stretch as $\Lambda(t)=1 + \dot{\Lambda}\,t$ with $\dot{\Lambda}=0.05\,s^{-1}$ for a time of $100~s$ followed by a constant stretch of $\Lambda(t)=5$ as shown in Fig~\ref{fig:stretch_rate}. The stress responses of all four developed models are presented together in Fig~\ref{fig:combined_model}, with the enlarged versions of the Poynting-Thompson model (PTM) and the Maxwell Model (MM) presented in Fig~\ref{fig:PTM_MM}. The material parameters used for this numerical example are given in Table~\ref{tab:material_parameter}(\text{A}). Fig.~\ref{fig:Viscous_strain} shows the evolution of the viscous strain over time for the Poynting-Thompson (PTM), Maxwell (MM) and the Zener model (ZM). As expected, one can observe a change in the slope of this curve in all three models around $t=100~s$ when the stretch rate reduces to zero. It is observed that the evolution of the viscous strain for the Maxwell and the Zener models coincide with each other. This is due to the fact that the evolution equations for the Maxwell and Zener model are the same as shown in\S~\ref{Sec:Maxwell_type_material}.

For the stress response, all four materials show a steady increase in stress for the first $100~s$ when the stretch $\Lambda$ is linearly increasing with time. Since the Maxwell and the Kelvin-Voigt materials can be obtained as limiting cases from the Poynting-Thompson and Zener solids respectively, as shown in \S~\ref{sec:limiting}, their respective responses within this time are noticeably similar. As the stretch is held constant after that, a stress relaxation behavior is prominently observed in the Maxwell, Zener, and the Poynting-Thompson model. Quite evidently, the Kelvin-Voigt model does not offer any noticeable relaxation and shows an elastic response under a zero stretch rate instead. On the other hand, the Zener model shows a stress relaxation that eventually aligns with the response of Kelvin-Voigt model. Similarly, the Poynting-Thompson and Maxwell model exhibit comparable stress responses during constant stretch rate.  While the Maxwell model eventually relaxes to an almost vanishing stress, Poynting-Thompson model relaxes to a constant value. Since the evolution equations for the Maxwell and the Zener model are identical (see Fig.~\ref{fig:Viscous_strain}), their stress relaxation behaviors are also similar albeit starting at and eventually relaxing to different values of stresses. 

\begin{figure}[h]
	\centering
	\begin{subfigure}{0.49\linewidth}
		\includegraphics[width=\linewidth]{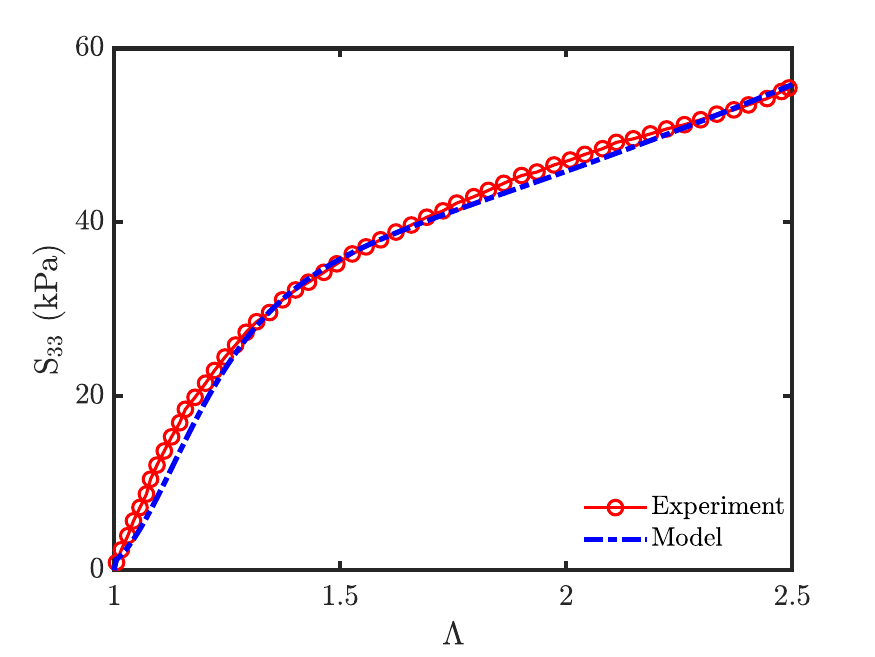}
		\caption{}
		\label{fig:PTM_Mokarram}
	\end{subfigure}
	\begin{subfigure}{0.49\linewidth}
		\includegraphics[width=\linewidth]{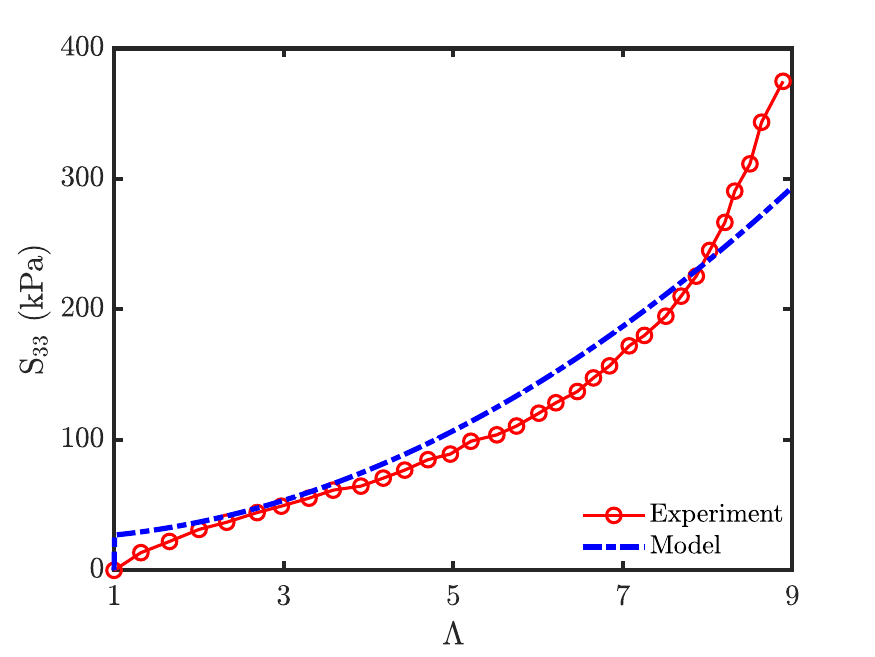}
		\caption{}
		\label{fig:PTM_Wang}
	\end{subfigure}
	\caption{Experimental verification of the Poynting-Thompson model: comparison with the experimental results of (a) Hossain~\textit{et al.}~(2012)~\cite{hossain2012experimental} up to a total stretch of $150\%$ and (b) Wang~\textit{et al.}~(2016)~\cite{wang2016modeling} up to a giant stretch of $800\%$.}
	\label{fig:VHB_experiment_validation}
\end{figure} 
\textbf{\textbullet Experimental verification of Poynting-Thompson solids:} Since the Zener and Maxwell model have long been explored, the developed Poynting-Thompson model is our primary model of interest. First, we verify this model with experimental results on the uniaxial stretching of polymers. Hossain~\textit{et al.}~(2012)~\cite{hossain2012experimental} and Wang~\textit{et al.}~(2016)~\cite{wang2016modeling} performed a uniaxial tension test on VHB 4910 polymer with stretch rates of $0.01~s^{-1}$, $0.03~s^{-1}$, $0.05~s^{-1}$ and $9\times10^{-5}~s^{-1}$, respectively.  Since the material is the same in both experiments, we keep the material parameter the same in our numerical modeling for both cases. These material parameters are presented in table~\ref{tab:material_parameter}(\text{B}). Here we use a linear stretch as $\Lambda (t)=1 + \dot{\Lambda}\,(t)$ with the stretch rates same as that ones used in the corresponding experiments. When comparing the response of our model with the experimental data of Hossain~\textit{et al.}~(2012)~\cite{hossain2012experimental} up to a total stretch of $150\%$, our model shows an excellent match with the experimental results, as shown in Fig~\ref{fig:PTM_Mokarram}. Similarly, the data from Wang~\textit{et al.}~(2016)~\cite{wang2016modeling} also show good alignment with our model up to a very large stretch of $700\%$, as indicated in Fig~\ref{fig:PTM_Wang}. Although the fit can be enhanced through an appropriate identification and optimization of the material parameters, it is outside of the scope of the current study and will be explored in the future.

\begin{figure}[h]
	\centering
    \begin{subfigure}{0.49\linewidth}
		\includegraphics[width=\linewidth]{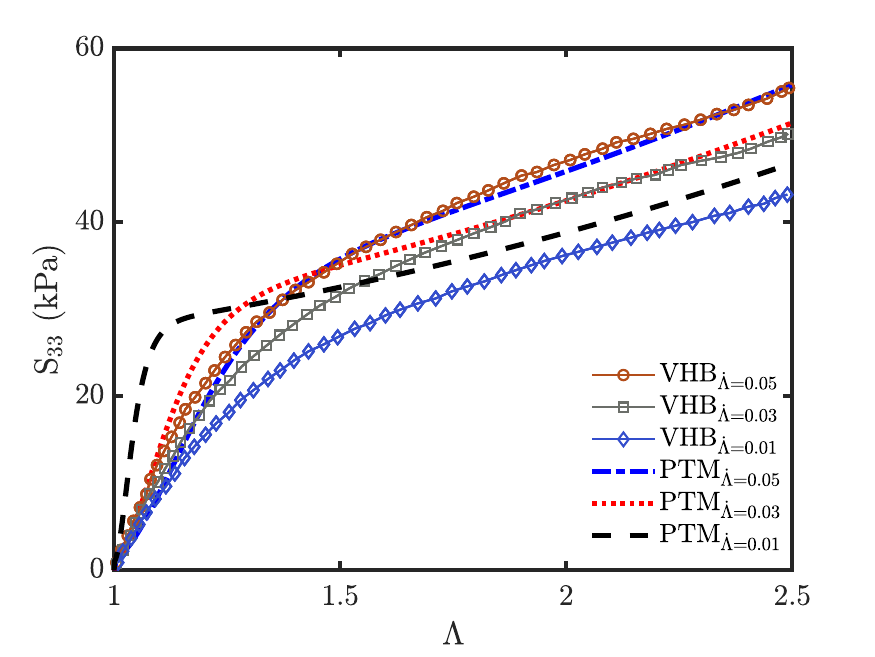}
		\caption{}
		\label{fig:PTM_Mokarram_different}
	\end{subfigure}
	\begin{subfigure}{0.49\linewidth}
		\includegraphics[width=\linewidth]{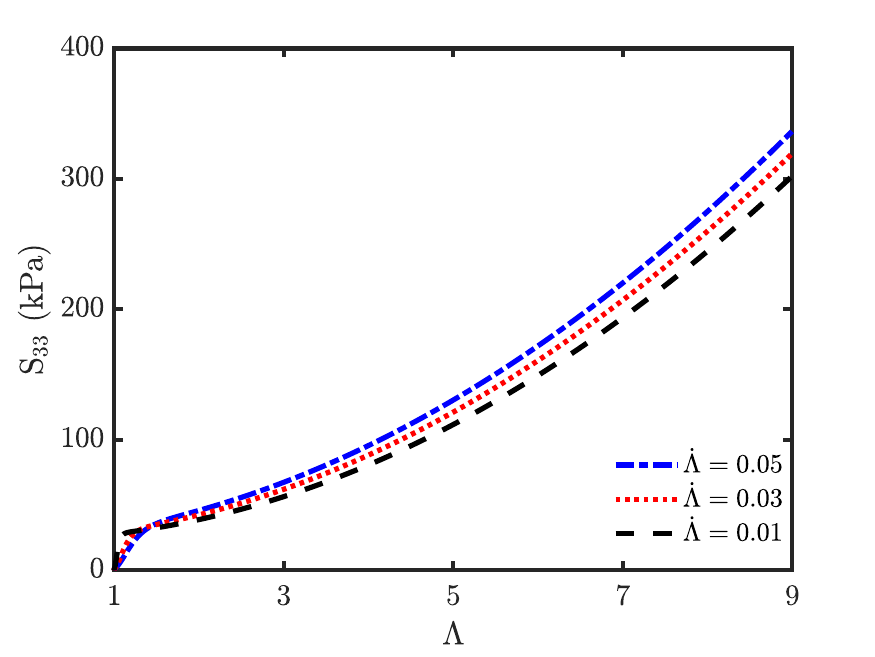}
		\caption{}
		\label{fig:PTM_different_stretch}
	\end{subfigure}
   \caption{Response of the Poynting-Thompson model with different stretch rates: (a) comparison with experimental data~\cite{hossain2012experimental} up to a total stretch of $250~\%$ and (b) effects of stretch rate up to a total stretch of $800~\%$.}
	\label{fig:VHB_experiment_validation}
\end{figure}

\textbf{\textbullet Response of the Poynting-Thompson solid at different stretch rates:} We now seek to understand the stress response of the Poynting-Thompson solid under various stretch rates. For this purpose, we select three different stretch rates of $0.05~s^{-1}$, $0.03~s^{-1}$ and $0.01~s^{-1}$ in order to compare with the experimental data of Hossain~\textit{et al}~(2012)~\cite{hossain2012experimental}. The results show that for $150\%$ stretch, the model shows an excellent fit at the stretch rate of $0.05~s^{-1}$. However, for the same extent of stretch at $0.03~s^{-1}$ and $0.01 ~s^{-1}$, the model aligns better at the later stages of the deformation as evident in Fig~\ref{fig:PTM_Mokarram_different}. At the range of total stretch up to $50\%$ similar curve, a deviation is observed between the model and experimental results which is more prominent at lower stretch rates such as at $0.01 ~s^{-1}$ as compared to $0.03 ~s^{-1}$. This initial deviation, however, does not affect the overall fit. When we consider a giant stretch of $ 800\%$, shown in Fig.~\ref{fig:PTM_different_stretch}, the model shows a similar trend in its response for all three stretch rates with necessary differences due to the rate-effects. It is important to note that this response fits good with the experimental data of Wang \textit{et al.}~(2016)~\cite{wang2016modeling} for a stretch rate of $9\times10^{-5}~s^{-1}$. It can be observed from Fig.~\ref{fig:PTM_different_stretch} that the response of the model is qualitatively similar to the experimental results of Treloar~(1976)~\cite{treloar1976mechanics} which was captured by the model of Boyce and Arruda~(2000)~\cite{arruda1993three}. Our model also shows a similar curve, an initial increase, followed by a plateau region where the stress increase is limited with the increase in stretch and thereafter, the model shows a sharp increase in stress as the stretch increases. For the lower stretch rates, as shown in Fig.~\ref{fig:PTM_Mokarram_different}, the plateau region is very prominent which causes it to deviate from the experimental curve at the initial stages of deformation. Nevertheless the deviation of our model from the experimental data is in the same order as that of Hossain~\textit{et al.}~(2012)~\cite{hossain2012experimental}. For their model, this deviation increases with the extent of stretch. On the other hand, although our model shows some deviation at the initial stage for low stretch rates, it performs significantly better at a very high extent of stretch as evidenced in Fig.~\ref{fig:PTM_Wang}.  
$\dot{\Lambda}$
\section{Concluding remarks}\label{Conclusion}
In this paper, we developed nonlinear viscoelastic models within the framework of evolving natural configuration. Based on the postulates of Singh and Paul~(2025)~\cite{singh2025}, we develop Maxwell and Kelvin-Voigt type material models using the strain and stress space formulations of the chosen framework, respectively. The highlight of the paper is the development of a novel large strain Poynting-Thompson solid model. We further show that a Maxwell and a Kelvin-Voigt model can be obtained as limiting cases from the developed Poynting-Thompson and Zener models, respectively. The integration algorithms for these models are developed and their efficacy has been shown by numerical implementation to the problem of a uniaxial stretch of polymers. The response of all four developed models for this problem has been discussed. The Poynting-Thompson model has been verified with experimental data and its response under different stretch rates has been demonstrated. For future direction, a full-scale finite element analysis can be developed for the proposed models. Here we have confined our attention to isotropic materials. The models can also be extended to other material symmetry groups by extending the set of invariants and modifying the functional form for the rate of dissipation function.     
\bibliographystyle{acm}

\end{document}